\documentclass[12pt]{article} 

\usepackage[utf8]{inputenc} 

\usepackage[margin=1.25in]{geometry}

\usepackage{graphicx} 

\usepackage{booktabs} 
\usepackage{array} 
\usepackage{verbatim} 
\usepackage{subfig} 

\usepackage{setspace}

\usepackage{rotating}
\usepackage{xy}

\usepackage[round]{natbib}

\usepackage{bbm}

\usepackage{color,hyperref}
\definecolor{darkblue}{rgb}{0.0,0.0,0.5}
\hypersetup{colorlinks,breaklinks,linkcolor=darkblue,urlcolor=darkblue,anchorcolor=darkblue,citecolor=darkblue}

\usepackage{amsbsy}
\usepackage{amsmath}
\usepackage{amsfonts}
\usepackage[flushleft]{threeparttable}

\usepackage{pifont}

\usepackage{color}

\usepackage{dsfont}

\usepackage{booktabs}
\usepackage{dcolumn}
\usepackage{tabularx}

 {\begin{list}{}%
         {\setlength{\leftmargin}{#1}}%
         \item[]%
 }
 {\end{list}}

\begin{document}

\title{\textbf{Boundary estimation in the regression-discontinuity design: Evidence for a merit- and need-based financial aid program}}


\author{Eugenio Felipe Merlano\thanks{Merlano: School of Economics and Finance,  Queen Mary University of London, Mile End Road, London, E1 4NS, U.K (e-mail: e.merlanolombana [at] surrey.ac.uk). I am deeply grateful to my advisors, Francesca Cornaglia and Claire Lim, for their guidance and continuous feedback throughout this project. I am thankful to Andrea Tesei, Sebastian Axbardand,  Karen Ortiz-Becerra, and seminar participants at various institutions for their comments. Any errors are my own.}}

\date{December, 2023}
\maketitle


\begin{abstract}
In the conventional regression-discontinuity (RD) design,  the probability that units receive a treatment changes discontinuously as a function of one covariate exceeding a threshold or cutoff point. This paper studies an extended RD design where assignment rules simultaneously involve two or more continuous covariates. We show that assignment rules with more than one variable allow the estimation of a more comprehensive set of treatment effects, relaxing in a research-driven style the local and sometimes limiting nature of univariate RD designs. We then propose a flexible nonparametric approach to estimate the multidimensional discontinuity by univariate local linear regression and compare its performance to existing methods. We present an empirical application to a large-scale and countrywide financial aid program for low-income students in Colombia. The program uses a merit-based (academic achievement) and need-based (wealth index) assignment rule to select students for the program. We show that our estimation strategy fully exploits the multidimensional assignment rule and reveals heterogeneous effects along the treatment boundaries.
\end{abstract}


\begin{spacing}{1.5}


\section{Introduction}

\begin{figure}[!h]
	\caption{Two-dimensional and three-dimensional discontinuities}
	\begin{minipage}[t]{0.3\textwidth} 
		\centering
		\includegraphics[scale = 0.4]{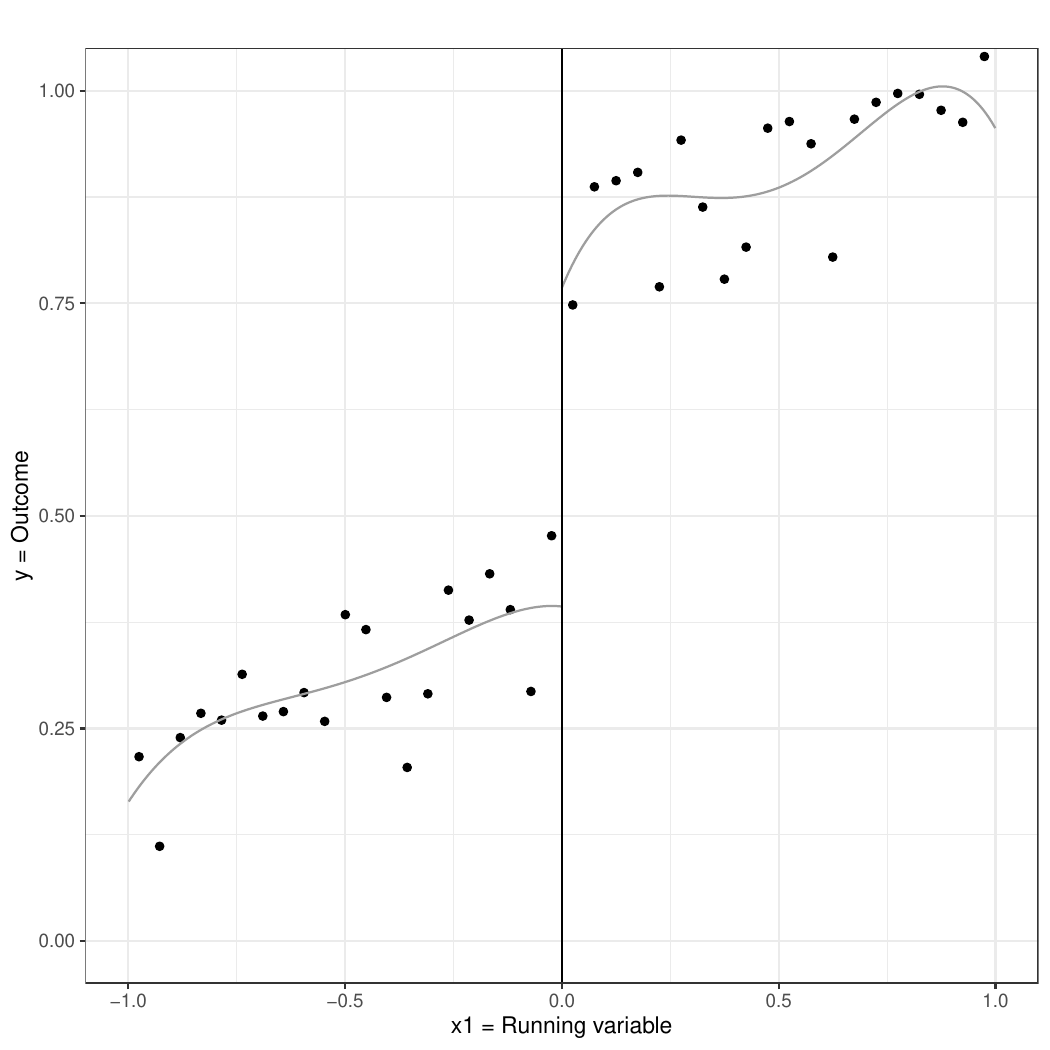}
		\caption*{\hspace{0.5cm} (a) Two-dimensional }
	\end{minipage}
	\hspace{2.3cm}
	\begin{minipage}[t]{0.3\textwidth}
		\centering
		\includegraphics[scale = 0.6]{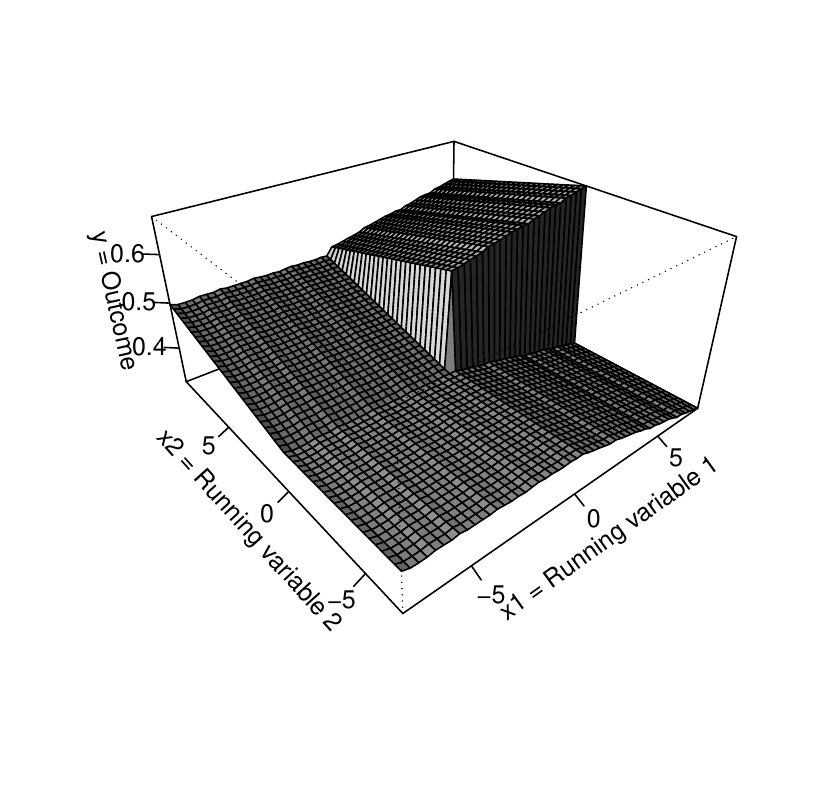}
		\caption*{\hspace{0.5cm} (b) Three-dimensional}
	\end{minipage}
	\caption*{\footnotesize \textbf{Note:} Figure \ref{fg:rdd} (a) displays the standard step-like discontinuity, where only one assignment variable (i.e., x-axis variable) is involved in the assignment mechanism. Each dot calculates a local sample mean over the support of the outcome variable $y$ and the two lines estimate a fourth-order global polynomial regression curve for control ($x_1<0$) and treated ($x_1 \geq 0$) units separately. Figure \ref{fg:rdd} (b) displays the case when two assignment variables (i.e., $x_1$ and $x_2$) generate a three-dimensional treatment boundary over the outcome $y$. The surface is drawn by a perspective plot of a discontinuous surface over the $x_1$–$x_2$ plane.}
	\label{fg:rdd}
\end{figure}

The regression-discontinuity (RD) design is one of the most popular and credible research designs in economics, political science, and education \citep[see, e.g.,][for recent reviews]{COOK2008636, Imbens2008,Skovron2015, Cattaneo_Escanciano}. The empirical method enables causal inference by exploiting the mechanism in which units (e.g., individuals, firms, classrooms) are assigned to treatment. In the standard RD design, the probability of receiving a particular treatment changes discontinuously as a function of one observable covariate exceeding a threshold. In this paper, we study more complex RD designs with assignment rules involving more than one variable simultaneously \citep[see, e.g.,][]{Papay2011,Zajonc2012}. For instance, in the simplest multidimensional case for continuous variables, a bivariate assignment rule generates a treatment boundary instead of a single step-like discontinuity. Figure \ref{fg:rdd} illustrates both the traditional two-dimensional step-like and the three-dimensional boundary-type discontinuities for simulated data. If treatment assignment involves more than one covariate at the same time (i.e., a vector of covariates with independent cutoffs), traditional RD estimates might not fully account for heterogeneous treatment effects along the treatment boundary.

Although the literature on RD designs has grown significantly in the last two decades \citep[see,][]{Cattaneo_Escanciano}, the study of RD designs with multiple assignment variables and the role of boundary effects is still at an early stage. Following \citet{Papay2011}, a few authors have proposed ways to analyzed and estimate RD with multiple running variables \citep[see, e.g.,][]{Reardon2012,Zajonc2012,Wong_etal_2013, Choi2018b,Choi2018}. The main focus of most procedures to date has been on dimensionality reduction of the multidimensional nature of the assignment rule to simplify the estimation of more complex treatment effects. \citet{Zajonc2012}, \citet{Cheng2020}, and \citet{Diaz2023} being notable exceptions.

In this paper, we focus on ways to exploit the multidimensional assignment rule to estimate potential heterogeneous treatment effects along the treatment boundaries for the case of two continuous running variables under the continuity-based framework\footnote{ An alternative is the local randomization framework which formulates the research design as a local randomized experiment near the cutoff \citep{Mattei2017}.} (treatment at the cutoff). We analyzed two of the main existing multidimensional methods for continuous running variables proposed in the literature. A semiparametric approach by \citet{Papay2011} and a nonparametric estimation by \citet{Zajonc2012}, and propose an alternative flexible nonparametric approach to estimate the multidimensional RDD by univariate local linear regression. The main purpose of this flexible approach is to simplify the estimation of boundary effects given the multidimensional nature of the estimates.  We also discuss some of the challenges faced by existing methods, focusing on sample size issues and the implication of the curse of dimensionality when defining neighbours around the cutoff in an RD setting.

We contribute to the growing literature on heterogeneous treatment effects in RD designs. Previous papers have focused on treatment effect heterogeneity in the response outcome \citep[e.g.,][]{Frandsen2012, Chiang2019}, observed individual characteristics \citep[e.g.,][]{Hsu2019,Reguly2021}, identification away from the threshold \citep[e.g.,][]{Dong2015,Angrist2015wanna}, and multiple cutoffs with a single running variable \citep[e.g.,][]{cattaneo21_multi}. This paper studies a different type of heterogeneity that occurs when multiple continuous running variables define selection into treatment, and that easily adapt to one or multiple cutoff points. We differ from other recent contributions in two ways. First, we study identification under the continuity framework for continuous running variables.  \citet{Diaz2023} study the case of multiple discrete and continuous running variables that builds on the local randomization framework.  Second, we propose estimation via standard local polynomial nonparametric methods with robust and bias-corrected inference \citep{Calonico2014}.  \citet{imbens_wager_19} propose an alternative minimax linear estimator for the regression discontinuity design that can be adapted to multiple running variables but requires additional tuning, and \citet{Cheng2020} suggests using thin plate regression splines that incorporate significant smoothing relative to our approach.

Some early applications of RDD models with multiple assignment variables were mainly found in education, such as various test scores (e.g., math and reading) selecting students into summer school and grade retention  \citep[e.g.,][]{Jacob04, Matsudaira2008}.  However, the scope of applications has recently expanded, and novel examples include international tax research \citep{Egger2015}, geographical analysis in urban economics \citep{Hidano2015}, general geographical boundaries \citep{Keele2015}, and financial aid targeting \citep{Londono19}.

We present an application of the multidimensional RD design to a merit- and need-based financial aid program for low-income students in Colombia. The main feature of the program is that it uses two main criteria to select students into the program: i) the student must score above a cutoff in the national standardized high-school exit examination, and ii) score below a specified cutoff in a wealth index for social programs. Our results show that exploiting the multidimensional nature of the treatment assignment for the program uncovers heterogeneous effects along the treatment boundaries. Compared to standard two-dimensional RD estimates, our nonparametric approach shows that estimated effects could be significantly smaller for some sub-population exposed to the policy. Furthermore, two-dimensional RD estimates could also involve endogenous averaging of significant and non-significant effects for different populations affected by the program. 

The rest of the paper is organized as follows. Section 2 and 3 describe the standard RD design and the extended RD with multiple assignment variables, respectively. Section 4 presents our flexible nonparametric approach to estimate the multidimensional discontinuity by univariate local linear regression. The approach simplifies and overcomes some of the limitations of existing methods. Section 5 describes the empirical application with further details on the program's design and the data used in the statistical analysis. Finally, Section 6 summarizes the results, and Section 7 closes with conclusions.


\section{The standard regression-discontinuity design}

Before considering the case of multiple assignment variables, we set the notation and review the main results for the conventional two-dimensional RD design to simplify the exposition in subsequent sections. Let $(Y_i,X_i)$ for $i = 1,\ldots,n$ be an $\mathbb{R}^2$-valued random sample, where $Y_i$ is an outcome variable and $X_i$ a predictor variable with density $f(x)$. The variable $X_i$ is also known in the RD literature as ``running'' or ``forcing'' variable. In the case of multiple assignment variables, we let $(Y_i,\mathbf{X}_i)$ for $i = 1,\ldots,n$ be an $\mathbb{R}^{d+1}$-valued random sample, where $\mathbf{X}_i$ is an $\mathbb{R}^{d}$-valued set of running variables with density $f_{d}(x)$.

In the standard RD design, the probability that unit $i$ receives treatment changes discontinuously as a function of $X_i$ exceeding a known threshold or cutoff $\bar{x}$, which we normalize hereafter to $\bar{x}=0$. We also define a treatment indicator as $T_i = \mathbf{1}(X_i \geq 0)$, where $\mathbf{1}(\cdot)$ is an indicator function that takes the value of one ($T_i = 1$) if the variable $X_i$ is greater or equal to zero, and zero ($T_i = 0$) otherwise. The fundamental difference between the two main classes of RD designs, the sharp and fuzzy RD, is that in the former, the probability of treatment assignment jumps from zero to one in a deterministic way when $X_i$ exceeds $\bar{x}$. In the fuzzy RD, the probability increases by less than one and other unobserved factors by the econometrician determine treatment assignment \citep{Hahn_etal_01}. Following \citet{Hahn_etal_01} and \citet{Calonico2014}, the parameter of interest in the RD design is the average treatment effect at the threshold $\tau = \mathbb{E}[Y_i(1) - Y_i(0) \mid X_i = \bar{x}]$, where $Y_i(1)$ and $Y_i(0)$ denote the potential outcome with and without treatment. Under the assumption of continuity at the threshold in the absence of treatment, a generalized $\tau$ can be nonparametrically identified as:

\begin{equation}\label{eq:tau}
\tau = \frac{\mu_{Y}^{+} - \mu_{Y}^{-}}{\mu_{T}^{+} - \mu_{T}^{-}} = \frac{\lim_{x \to 0^+}\mu_{Y}(x) - \lim_{x \to 0^-}\mu_{Y}(x)}{\lim_{x \to 0^+}\mu_{T}(x) - \lim_{x \to 0^-}\mu_{T}(x)},
\end{equation} \

\noindent where $\mu_{Y}(x)=\mathbb{E}[Y_i \mid X_i = x]$ and $\mu_{T}(x)=\mathbb{E}[T_i \mid X_i = x]$. For the sharp RD design, the denominator in (\ref{eq:tau}) becomes equal to one given that the  $\lim_{x \to 0^+}\mathbb{E}[T_i \mid X_i = x]=1$ and $\lim_{x \to 0^-}\mathbb{E}[T_i \mid X_i = x]=0$, and $\tau$ can be redefined as $\tau_{SRD}$. For the fuzzy RD, the presence of non-compliers makes the denominator different from 1 and $\tau = \tau_{FRD}$. Given its boundary properties, the current standard to consistently estimate the four limits in (\ref{eq:tau}) is to use local polynomial regression estimators \citep[see, e.g.,][]{Fan1996,Hahn_etal_01}.\footnote{It is worth mentioning that identification, estimation, and inference methods for RD designs are currently expanding (see, e.g., \citet{imbens_wager_19} and \citet{eckles_20} for recent contributions in the area).}


\section{RD designs with multiple assignment variables}

The literature on the standard RD design has consolidated significantly in the last couple of decades with significant improvements in identification and estimation \citep[e.g.,][]{Hahn_etal_01}, optimal bandwidth selection \citep[e.g.,][]{Imbens2011}, and inference \citep[e.g.,][]{Calonico2014}. However, contributions in the case of multidimensional assignment rules have been more limited (see, e.g., \citet{Cattaneo2022rev} for a recent and comprehensive review).

The few exceptions that have discussed the case of assignment rules with two or more variables have almost exclusively focused on dimensionality reduction of the RD design to simplify the estimation of treatment effects as a standard two-dimensional RD design.\footnote{\citet{Papay2011}, \citet{Zajonc2012}, and \citet{imbens_wager_19} being notable exceptions discussed later in this paper.} For instance, \citet{Reardon2012} discuss RD designs with multiple assignment variables in which the estimation methods focus on either averaging treatment effects for the different assignment variables or local average treatment effects through boundary-specific discontinuities. In both cases, the focus is on reducing the multidimensional feature of the RD design. The authors also distinguish the case of assignment to one treatment condition by multiple assignment variables and the case of assignment to various treatments by multiple assignment variables. In this paper, we focus on the assignment to one treatment/control condition and consider the extension to different treatment as a special case.

In the same line, \citet{Wong_etal_2013} present a Monte Carlo simulation of four estimation procedures that the authors call: the i) centering, ii) instrumental variable (IV), iii) univariate or conditional, and iv) frontier approach.\footnote{Unfortunately, the literature has not reached consensus in the names attributed to the estimation methods, and names are mainly paper-specific.} Again, three out of the four methods use dimensionality reduction of the assignment rule, with the exception fundamentally intended to average the boundary effect as an average treatment parameter. The three methods discussed by the authors that use standard two-dimensional RD procedures (i.e., centering, IV, and univariate approach) are probably the most common in empirical studies. In the centering approach, all assignment variables are centered at their thresholds $\mathbf{X}_i^c = [X_{1,i} - \bar{x}_1, \ldots, X_{d,i} - \bar{x}_d]'$, and a new assignment variable is defined for each unit as its minimum centered score $X_i^* = \text{min}\{ X_{1,i} - \bar{x}_1, \ldots, X_{d,i} - \bar{x}_d \}$. The new assignment variables $X_i^*$ is used as the unique running variables in a conventional RD estimation (see, e.g., \citet{Zimmer07} for an application). In the IV approach, one of the assignment variables is used as an instrument for treatment assignment with the second running variable facing treatment misallocation \citep[see, e.g.,][]{Wong_etal_2013}. Finally, in the univariate analysis, each running variable is treated independently in a traditional two-dimensional RD and estimations are conditional on the subsample of units assigned to treatment by the second assignment variables. \citet{Londono19} apply this strategy to an impact evaluation of the financial aid program in Colombia that we study later in this paper known as ``Ser Pilo Paga''.

Finally, \citet{Choi2018b} and \citet{Choi2018} focus on multiple running variables allowing partial effects, which is the case when only one score crossing a cutoff affects an outcome of interest. Given that we focus on complete boundary estimations and we do not face partial effects in our empirical application, we do not discuss Choi and Lee's results in details here.

It is essential to clarify that we do not regard dimensionality reduction methods as not desirable. As shown and discussed by existing papers, these methods have helped overcome some of the challenges of multidimensional assignment rules. However, we show that researchers and practitioners might also benefit from exploiting multivariate assignment rules to study potential heterogeneous effects or gain further insight into the nature of the treatment effect under consideration. Ultimately, the two methods can be complementary, and the multidimensional methods can retrieve standard RD estimates.

In the next sections, we first discuss the two main contributions that have exploited multidimensional assignment mechanisms in RDD, and then, we propose an alternative new estimation approach.

\subsection{Semiparametric approach}

Most of the literature on multidimensional RD designs started with \citet{Papay2011}. The authors consider an extension of the conventional RD design to the case of two assignment variables when various criteria determine placement into different treatment conditions. The appeal of the bivariate assignment rule is not exclusively a way to circumvent further estimation challenges but also a desirable graphical feature of RD designs that is lost in higher dimensions. As shown in Figure  \ref{fg:rdd} (b), a bivariate rule causes a three-dimensional boundary. More importantly, assignment rules in higher dimensions are likely to induce the curse of dimensionality provided that RD estimations rely on observation within a bandwidth or neighbourhood around the threshold value. In this paper, we focus on the two-dimensional assignment rule provided the nature of our empirical application and consider higher dimensions as extensions.

In what follows, we refer to \citet{Papay2011}'s contribution as the \textit{semiparametric approach} given that the method consists of a regression model with 16 parameters that fits all the possible interactions of the two running variables $X_{1,i}$ and $X_{2,i}$, and the two binary treatment indicators $T_{1,i}$ and $T_{2,i}$. Although this is a parametric specification, the authors select the bandwidth nonparametrically through cross-validation. Analogously to the two-dimensional case, $X_{1,i}$ and $X_{2,i}$ form the $\mathbb{R}^{2}$-valued set of forcing variables $\mathbf{X}_i$ centered on their respective thresholds. $T_{j,i}$ for $j \in \{1,2\}$ and $i = 1,\ldots,n$ describes the two treatment conditions given the known threshold $\bar{x}_j$ as $T_{j,i} = \mathbf{1}(X_{j,i} \geq \bar{x}_j)$ for each assignment variable.

As described by \citet{Papay2011}, the model fits four planes in three dimensions with an intercept at both thresholds ($\bar{x}_1$, $\bar{x}_2$). Each plane lies in one of the four possible regions defined by the treatment interaction of $T_1$ and $T_2$.\footnote{ The 2-by-2 quadrants are defined as follows: i) $T_1 = 0 \cap T_2 = 0$, ii) $T_1 = 0 \cap T_2 = 1$, iii) $T_1 = 1 \cap T_2 = 0$, and iv) $T_1 = 1 \cap T_2 = 1$.} This approach studies the discontinuities of the four planes at the edges. As in the standard RD design, estimations in the bivariate assignment rule rely on observations within a bandwidth around the threshold boundary. Following \citet{Imbens2008}, the authors generalize the two-dimensional cross-validation criterion to the case of two assignment variables.\footnote{ As discussed in Section \ref{mer}, to make estimations comparable with other methods in this paper, we use \citet{Calonico2014} optimal bandwidth selector instead.}

As noted by \citet{Zajonc2012}, unlike the conventional RD for which there is one optimal bandwidth, the optimal bandwidth varies along the boundary. \citet{Papay2011} and \citet{Zajonc2012} simplify the estimation procedure by selecting a single pair of bandwidths. Although all empirical methods to date use a unique optimal bandwidth, the method we propose in Section \ref{mer} relaxes this restriction and allows for a flexible optimal bandwidth selection along the boundaries following \citet{Calonico2014} optimal bandwidth selector.

We do not discuss confidence intervals for effects along each boundary for this procedure, given that there is no such treatment in the original paper by \citet{Papay2014}. Nevertheless, when estimating regression like the ones develop by Papay and coauthors, we consider conventional inference procedures for point estimates. We highlight the performance and discuss further benefits and drawbacks of this approach later in the empirical application.

\subsection{Nonparametric approach}

In a subsequent approach, \citet{Zajonc2012} extends \citet{Hahn_etal_01} seminal paper on identification by local linear estimations to the case of two or more running variables. The nonparametric approach as we call it hereafter identifies both the conditional treatment effect at any point on the treatment boundary as well as the average treatment effect over the entire boundary. The author also derives an optimal data-dependent bandwidth selection procedure in line with \citet{Imbens2011} plug-in estimator.

Under a boundary positivity assumption and a continuity assumption for the multidimensional case, \citet{Zajonc2012} shows that a fuzzy boundary RD (FBRD) estimate $\tau_{FBRD}(\mathbf{x})$ can be nonparametrically identified as:

\begin{equation}\label{eq:fbrd}
\tau_{FBRD}(\mathbf{x}) = \frac{\mu_{Y}^{+}(\mathbf{x}) - \mu_{Y}^{-}(\mathbf{x})}{\mu_{T}^{+}(\mathbf{x}) - \mu_{T}^{-}(\mathbf{x})} = \frac{\lim_{\epsilon \to 0}\mu_{Y}(\mathbf{x}) - \lim_{\epsilon \to 0}\mu_{Y}(\mathbf{x})}{\lim_{\epsilon \to 0}\mu_{T}(\mathbf{x}) - \lim_{\epsilon \to 0}\mu_{T}(\mathbf{x})},
\end{equation}\

\noindent where $\mu_{Y}(\mathbf{x})=\mathbb{E}[Y_i \mid \mathbf{X}_i \in N^{+/-}_{\epsilon}(\mathbf{x}_i)]$, $\mu_{T}(\mathbf{x})=\mathbb{E}[T_i \mid \mathbf{X}_i \in N^{+/-}_{\epsilon}(\mathbf{x}_i)]$, and $N_{\epsilon}(\mathbf{x}_i) \equiv \{\mathbf{X} \in \boldsymbol{\mathcal{X}}: (\mathbf{X} - \mathbf{x})'(\mathbf{X} - \mathbf{x}) < \epsilon^2 \}$. $N_{\epsilon}(\mathbf{x}_i)$ denotes the $\epsilon$-neighborhood around $\mathbf{x}$ that contains all points $\mathbf{X}$ within a sphere of radius $\epsilon$ around $\mathbf{x}$. The positive and negative superscript denote the points in the $\epsilon$-neighborhood around $\mathbf{x}$ that receive treatment (i.e., $N^{+}_{\epsilon}(\mathbf{x}_i) \equiv N_{\epsilon}(\mathbf{x}_i) \cap \mathbb{T}$) and the observations in the control group (i.e., $N^{-}_{\epsilon}(\mathbf{x}_i) \equiv N_{\epsilon}(\mathbf{x}_i) \cap \mathbb{T}^c$). Note that it is  explicit the dependence of the effect on a point along the boundary $\mathbf{x}$. The sharp BRD $\tau_{SBRD}(\mathbf{x})$ in the absence of non-compliers is simply the numerator on equation (\ref{eq:fbrd}) $\tau_{SBRD}(\mathbf{x}) = \mu_{Y}^{+}(\mathbf{x}) - \mu_{Y}^{-}(\mathbf{x}) = \lim_{\epsilon \to 0}\mu_{Y}(\mathbf{x}) - \lim_{\epsilon \to 0}\mu_{Y}(\mathbf{x})$.

Following \citet{Hahn_etal_01}, \citet{Zajonc2012} consistently estimates the limits in equation (\ref{eq:fbrd}) by an equivalent multidimensional local linear regression estimators \citep[see, e.g.,][]{Fan1996}. These estimators are extensions of the multivariate locally weighted least squares regression by \citet{Ruppert94}. Assuming the bandwidth matrices for treated and untreated units needed for the local linear regressions are selected such that the asymptotic bias disappears, valid inference could be based on conventional robust standard errors.

Although appealing, the limitations of the nonparametric approach include several practical difficulties identified by the author in the choice of the bandwidth matrices. For instance, there is no closed-form solution for the optimal choice of the bandwidth matrix, dependence on the boundary's precise shape, and the estimation of the Hessian matrices for $\hat{\mu}_{Y}^{+}(\mathbf{x}) = m_0(\mathbf{x})$ and $\hat{\mu}_{Y}^{-}(\mathbf{x})=m_1(\mathbf{x})$ \citep[see,][]{Zajonc2012}. Moreover, in our empirical application, we identify further constraints such as the selection of a fixed bandwidth along the different boundaries. This bandwidth is especially problematic when estimations rely on running variables with different metric and scale, moderate sample size, and discrete outcome variables. Finally, the method implies a demanding implementation for practitioners and researchers, given the lack of statistical packages to perform the onerous algorithm smoothly. In a recent contribution, \citet{cattaneo2020stata} introduced the first statistical package for RD designs with multiple cutoffs or multiple scores. Nevertheless, the authors do not focus on the case of complete estimation of the treatment boundary as we do in this paper.

In the next section, we propose an alternative and flexible nonparametric approach that simplifies the fully multidimensional RD method. We perform the estimation by recovering the complete treatment boundary with simple two-dimensional local linear regression along the cutoff boundary.  The critical point of this alternative approach is to make more flexible for practitioners the estimation of boundary treatment effects and to overcome some of the current limitations of the methods discussed so far.


\section{An alternative nonparametric approach} \label{mer}

The idea of an alternative nonparametric approach is twofold. First, we show that existing methods do not fully adapt to the nature of the data studied in the empirical application of this paper. Second, more flexible and simple methods might contribute to expand the range of application using boundary effect in the presence of multiple assignment rules. We apply local liner regression in a two-dimensional RD-fashion along the different boundaries.  In this way, we make more flexible the implicit parametric specifications in methods like \cite{Papay2011}, and reduce the challenges imposed by the fully multidimensional nature of estimations like \citet{Zajonc2012}. Furthermore, our estimation procedure might be useful for empirical applications that face small sample size restrictions or running variables that do not share the same scale or metric.

To formally define the treatment boundary $\mathbb{B}$, we build on \cite{Zajonc2012} notation by defining a generic ``assignment rule,'' $\delta(\mathbf{x}) \colon \mathcal{X} \mapsto \{0,1\}$, as a function that maps units with covariates  $\mathbf{X}=\mathbf{x}$ to treatment assignment $T$.\footnote{ Hereafter, we denote with lowercase any specific pair of points $(x_1, x_2) = \mathbf{x}$ where $\mathbf{x} \in \mathbf{X}$.} For example, the scatter plot in Figure \ref{fg:brdd} (a) illustrates in the space of running variables $(X_1,X_2)$ a discontinuous assignment rule $\delta(\mathbf{x})$ for each point that takes the form $\delta(x_1,x_2)=\mathds{1}\{x_1 \geq 0, x_2 \geq 0\}$. The treatment assignment for any unit $(X_{1,i}, X_{2,i})$ is defined as $Z = \delta(\mathbf{X})$.

For example, in the empirical application that we study in Section \ref{emp}, $X_1$ represents a standardized test score, and $X_2$ represents a poverty index. High school students are eligible for financial aid if they have a test score above a cutoff of 310 points ($\bar{x}_1$) out of 500, and simultaneously have a poverty index over a second cutoff of 57.21 ($\bar{x}_2$) out of 100 points. Anyone with either score below the corresponding threshold is not eligible for the program.

Given the generic assignment rule $\delta(\mathbf{x})$, we can define a treatment assignment set $\mathbb{T}$ (and by extension a complement or ``control'' set $\mathbb{T}^c$) as $\mathbb{T} \equiv \{\mathbf{x} \in \mathcal{X} \colon \delta(\mathbf{x}) = 1   \}$. The assignment boundary $\mathbb{B}$ is defined as $\mathbb{B} \equiv \mathtt{bd}(\mathbb{T}) \equiv \bar{\mathbb{T}} \cap \bar{\mathbb{T}}^c $ where overbars denote the closure of the set. A point $\mathbf{x}$ is in the assignment boundary if the point contains observations in both, the treatment set $\mathbb{T}$ and the control set $\mathbb{T}^c$. Note that in a bivariate assignment to treatment rule (i.e., $\mathbf{X}_i \in \mathbb{R}^2)$, being on one of the treatment boundaries implies keeping fixed the value one of the two running variables. Data points in the non-fixed running variables are used to move along that boundary  as shown in Figure \ref{fg:brdd} (b).

To illustrate the estimation procedure, we follow the hypothetical discontinuities in Figure \ref{fg:brdd} (b), where the two boundaries of interest are:\footnote{ This is also the kind of discontinuities that we analyze in the empirical application in Section \ref{emp}.}

\begin{equation}
\mathbb{B}_1 \equiv \{ {\mathbf{x} \in \mathbf{X}: x_1 = 0 \cap x_2 \geq 0 } \}
\end{equation}

\noindent and

\begin{equation}
\mathbb{B}_2 \equiv \{ {\mathbf{x} \in \mathbf{X}: x_1 \geq 0 \cap x_2 = 0} \}.
\end{equation}\

Figure \ref{fg:brdd} also illustrates the idea of (linear) heterogenous treatment effects along the boundary.   In the case of $\mathbb{B}_1$, the discontinuous jump or treatment effect for units with value closer to $x_2 = 0$ is larger than the discontinuity for those observations away from $x_2 = 0$. On the other hand, the reverse is true for boundary $\mathbb{B}_2$ and $x_1 = 0$. The procedure to estimate the discontinuities along $\mathbb{B}_1$ and $\mathbb{B}_2$ consist of running local linear regressions for each point along each boundary. For instance, for a given point $\mathbf{x} \in \mathbb{B}_j$ for $j \in \{1,2\}$, the sharp RD $\tau_{SBRD \lvert \mathbb{B}_j }(\mathbf{x})$ can be identified as follow:

\begin{figure}[h]
\centering
\caption{Simulated assignment rule and boundary discontinuity}
	\begin{minipage}[t]{0.45\textwidth} 
		\includegraphics[scale=.56]{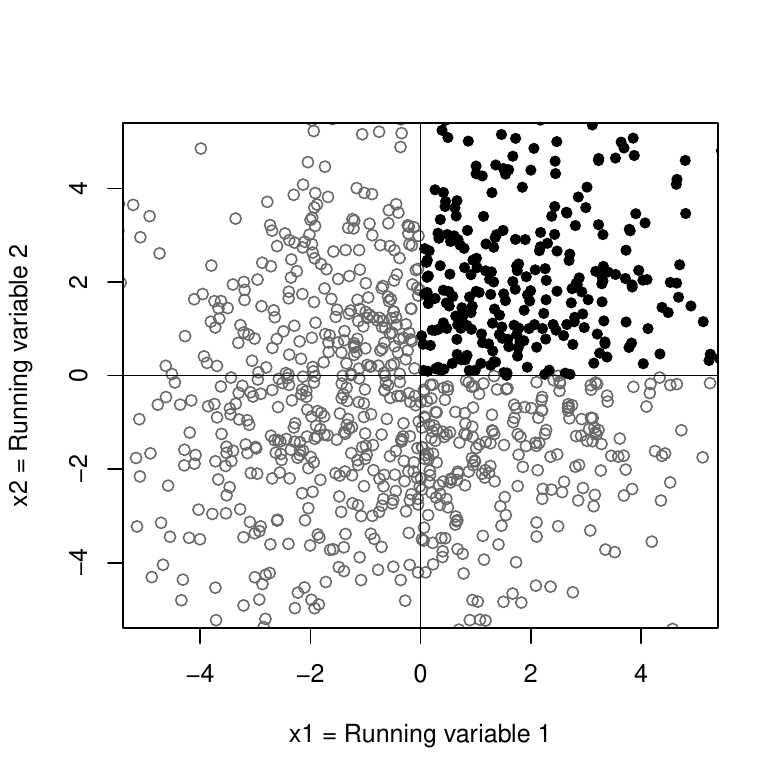}
		\caption*{\hspace{0.5cm} (a) Assignment rule}
	\end{minipage}
	\begin{minipage}[t]{0.45\textwidth}
		\includegraphics[scale=.6]{figures/brdd_1.pdf}
		\caption*{\hspace{0.5cm} (b) Simulated boundary}
	\end{minipage}
\caption*{\footnotesize \textbf{Note:} Figure \ref{fg:brdd} (a) illustrates a discontinuous assignment rule $\delta(\mathbf{x})$ that takes the form $\delta(x_1,x_2)=\mathbf{1}\{x_1 \geq 0, x_2 \geq 0\}$. Figure \ref{fg:brdd} (b) plots the simulated three-dimensional boundary. The surface is drawn by a perspective plot of a discontinuous surface over the $x_1$–$x_2$ plane.}
\label{fg:brdd}
\end{figure} \

\begin{equation}\label{eq:sbrd_new}
\tau_{SBRD \lvert \mathbb{B}_j}(\mathbf{x}) = \mu_{Y \lvert \mathbb{B}_j}^{+}(\mathbf{x}) - \mu_{Y \lvert \mathbb{B}_j}^{-}(\mathbf{x}) = \lim_{x_j \to 0^+}\mu_{Y \lvert \mathbb{B}_j}(\mathbf{x}) - \lim_{x_j \to 0^-}\mu_{Y \lvert \mathbb{B}_j}(\mathbf{x}),
\end{equation}\

\noindent where $\mu_{Y \lvert \mathbb{B}_j}(\mathbf{x}) = \mathbb{E}[Y_i \mid \mathbf{X}_i \in \mathbb{B}_j(\mathbf{x})]$ for $j \in \{1,2\}$. There are two main differences relative to the standard RD estimates like the one in equation (\ref{eq:tau}). First, it is explicit the dependence of $\tau_{SBRD \lvert \mathbb{B}_j}(\mathbf{x})$ on a point $\mathbf{x}$. The ability to estimate a wider set of parameter throughout the different boundaries is the main appealing of the multidimensional approach. Second, $\tau_{SBRD \lvert \mathbb{B}_j}(\mathbf{x})$ is at the same time conditional on one of the boundaries $\mathbb{B}_j$. As it is visible in Figure \ref{fg:brdd} (b), when observations lie on boundary $\mathbb{B}_1$, the variable that induces a discontinuous jump is $X_1$ with both treatment and control observations to either side of the edge and values of $X_2$ set to be greater than zero. Analogously, when points lie on $\mathbb{B}_2$, $X_2$ is the variable that causes the RD-type discontinuity with values of $X_1$ set to be greater than zero. We capture the boundary-specific running variable in equation (\ref{eq:sbrd_new}) with the limits over $x_j$ for $j \in \{1,2\}$. 

Similarly, the equivalent fuzzy RD $\tau_{FBRD \lvert \mathbb{B}_j}(\mathbf{x})$ can be identified as:

\begin{equation}\label{eq:fbrd_new}
\tau_{FBRD \lvert \mathbb{B}_j}(\mathbf{x}) = \frac{\mu_{Y \lvert \mathbb{B}_j}^{+}(\mathbf{x}) - \mu_{Y \lvert \mathbb{B}_j}^{-}(\mathbf{x})}{\mu_{T \lvert \mathbb{B}_j}^{+}(\mathbf{x}) - \mu_{T \lvert \mathbb{B}_j}^{-}(\mathbf{x})} = \frac{\lim_{x_j \to 0^+}\mu_{Y \lvert \mathbb{B}_j}(\mathbf{x}) - \lim_{x_j \to 0^-}\mu_{Y \lvert \mathbb{B}_j}(\mathbf{x})}{\lim_{x_j \to 0^+}\mu_{T \lvert \mathbb{B}_j}(\mathbf{x}) - \lim_{x_j \to 0^-}\mu_{T \lvert \mathbb{B}_j}(\mathbf{x})},
\end{equation}\

\noindent where $\mu_{T \lvert \mathbb{B}_j}(\mathbf{x}) = \mathbb{E}[T_i \mid \mathbf{X}_i \in \mathbb{B}_j(\mathbf{x})]$ for $j \in \{1,2\}$. As in the conventional RD, we consistently estimate the limits in equations (\ref{eq:sbrd_new}) and (\ref{eq:fbrd_new}) by local linear regressions. We follow \citet{Calonico2014} robust bias-corrected estimation and inference procedures which are available in leading statistical software like Stata and R \citep[see,][]{calonico14stata,calonico14r}.

There are, however, two further issues to consider. First, the selection of points along the boundaries $\mathbb{B}_1$ and $\mathbb{B}_2$ at which we perform the estimations. Second, the specific procedure to obtain enough mass of points (observations) at each point when using empirical data with limited sample size. For the former, we suggest selecting a grid of points by using percentiles of the distribution at each boundary.  In practice, any point on the boundary is valid in terms of identification as long as there is mass to perform the estimation.\footnote{ This is particularly helpful to make estimations flexible when the sample size is small or when the distribution of a running variable is highly skewed.} To address the latter consideration, we select a fraction $F_{\eta}$ of the data around each point. The need to use a portion of neighbour points comes from the fact that it is unlikely in empirical settings to obtain enough observations at each point along the edges to perform the two-dimensional RD estimations with one of the running variables.  Since identification comes from comparing units at each side of the running variables and not from the fraction of points itself, the role of $F_{\eta}$ is mainly about soothing the estimates along the frontier.  In our empirical application, we use a moving window along the boundary that takes 10 percent of the data points of the distribution.  We also show that results are robust if we fix the observation window (no overlap) or modify the selected fraction of the data (e.g., 5 or 20 per cent).  Using a fix fraction of observations also keeps the sample size for each point estimate relatively constant and makes discontinuities effects more comparable. Finally, given that we select the points to perform the estimation using percentiles of the running variables, we move by increments of 1 percent to estimate the complete discontinuity throughout the boundary.\footnote{ In our empirical application, although SABER 11 test scores range continuously from 0 to 500, reported scores are rounded to the nearest integer by the evaluation authority. To keep variables in their original format, we treat test scores as a continuous variable implying some repeated percentile for this variable. If estimations lie in the same window in the empirical application, we ignore them as they provide the exact value as a neighboring estimation.} 

Once the grid and the fraction of points is selected, we perform local linear regressions at each selected $\mathbf{x}$ along the boundaries \citep[see,][]{Hahn_etal_01,Calonico2014}.\footnote{ Even though this procedure might appear like conventional heterogeneous treatment effects on one covariate, in this case, the variables at hand happen to be running variables in a regression discontinuity framework, which makes identification, estimation, and inference distinct.} For instance, the two limits for the sharp case (\ref{eq:sbrd_new}) are the difference in intercepts of the following two first-order local polynomial regressions: 

	\begin{equation} \label{eq:em1}
	\small
	\begin{aligned}
	\hat{\mu}_{Y \lvert \mathbb{B}_j}^{+} (\mathbf{x}) = \underset{\beta_0(\mathbf{x})\in \mathbb{R}}{\mathrm{arg \hspace{0.1cm} min}} \sum_{i=1}^{n}\mathbf{1}(X_{j,i} \geq 0)(Y_i - \beta_0(\mathbf{x}) + \beta_1(\mathbf{x})(X_{j,i}-x_j))^2 K\left( \dfrac{X_{j,i} - x_j}{h^{*}_{\mathbf{x}}} \right)
	\end{aligned}
	\end{equation}
	
\noindent and

	\begin{equation} \label{eq:em2}
	\small
	\begin{aligned}
	\hat{\mu}_{Y \lvert \mathbb{B}_j}^{-} (\mathbf{x}) = \underset{\beta_0(\mathbf{x})\in \mathbb{R}}{\mathrm{arg \hspace{0.1cm} min}} \sum_{i=1}^{n}\mathbf{1}(X_{j,i} < 0)(Y_i - \beta_0(\mathbf{x}) + \beta_1(\mathbf{x})(X_{j,i}-x_j))^2 K\left( \dfrac{X_{j,i} - x_j}{h^{*}_{\mathbf{x}}} \right)
	\end{aligned},
	\end{equation}\

\noindent for $j \in \{1,2\}$ and each estimation using the pre-specified fraction of points. In equations (\ref{eq:em1}) and (\ref{eq:em2}), $h^{*}_{\mathbf{x}}$ indicates the two-dimensional optimal bandwidth employed in traditional RD designs. To select $h^{*}_{\mathbf{x}}$ we follow \citet{Calonico2014} MSE-optimal bandwidth selection. One key difference concerning \citet{Zajonc2012} is that our bandwidth is flexibly estimated at each point $\mathbf{x}$ on the treatment boundary rather fixing a single optimal bandwidth. \citet{Zajonc2012} selects the bandwidth for both boundaries as the smallest plug-in bandwidth from an evenly spaced grid of points. As we show in Section \ref{results}, for our empirical application, forcing a single bandwidth is potentially too restrictive given the differences in scale and metric of the two running variables that we use (i.e., test score and wealth index).\footnote{ The constraint is no adequately alleviated by forcing the variables onto the same scale as proposed by \citet{Zajonc2012}.} We discussed further implications of the fixed-bandwidth method when comparing the performance of the different estimation procedures.

In the next sections, we first describe the empirical application to a merit- and need-based financial aid program for low-income students in Colombia, and then discuss the statistical performance of the methods discussed so far.


\section{Empirical application} \label{emp}

We test the performance of the boundary RD design with an empirical application where we estimate the impact of a merit-based scholarship program for low-socioeconomic status (SES) in Colombia (see, \citet{DNP_2016} and \citet{Londono19} for a comprehensive description of the program). The financial aid program called ``Ser Pilo Paga'' (hereafter, SPP) was introduced during four years (2014-2017), and it is known as one of the leading higher education policies in recent year in Colombia. The program's main feature is that it uses two main criteria to select students into the program.  First, students must score above a cutoff in the standardized high-school exit examination,  known as ``SABER 11''.  Test scores serve as a merit-based running variable in the RD jargon.  Second, eligible students are required to score below a cutoff in a wealth index for social assistance know as ``SISBEN''.  This need-based index works as the second running variable. The selection mechanism fits the idea of a two-dimensional assignment rule perfectly and constitutes a novel application of the multidimensional RD design.

In the next sections, we describe details of the program and the data we use throughout the empirical analysis.


\subsection{Program design}

In October 2014, the Colombian government announced the most extensive higher education merit- and need-based financial aid program in the history of the country. The program offered since 2014 and for four yearly cohorts (2014-2017) a fully forgivable loan or scholarship upon graduation to low-income students with the highest scores in the national standardized high-school exam in Colombia (henceforth SABER 11). The target was to enroll 10,000 low-income students annually to complete 40,000 students in the top public and private universities of the country.\footnote{ This was roughly the same number of enrolled students, around 43 thousand, in the largest and national-wide public university of the country (i.e., National University of Colombia) at the time of the announcement.} The first criterion for eligibility was that students must score above 1.4 standard deviations from the mean in the national exam. This threshold is equivalent to a score above 310 out of the 500 maximum score. As describe by \citet{Londono19}, the test is generally alike to the SAT in the United States with minor differences.  For instance, the exam is an official national requirement for high school graduation, and it is widely used for admissions to post-secondary education in Colombia.  The 310 cutoff selects approximately the top 9 percent of the distribution.

The second criterion was that pupils must also come from a low-income household in terms of the national wealth index for social programs known as SISBEN III\footnote{ \textit{Sistema de Selección de Beneficiarios para Programs Sociales} for its Spanish name.}. The wealth index is a proxy-means testing instrument with continuous scores from 0 to 100 where higher values indicating higher wealth \citep[see, e.g.,][]{sisben}. The student's family score had to be below a prespecified cutoff that varied with geographical location. For instance, eligibility thresholds differed as follow: scores below 57.21 for students in the 14 metropolitan areas (i.e., main cities in the country), below 56.32 for other urban areas, and below 40.75 for rural areas. 

Although this running variable faces multiple cutoffs, we would initially proceed with the conventional simplification of normalizing the score variable at zero and using the normalized score to estimate a pooled RD design \citep{cattaneo16_multi}. Nonetheless, as discussed by \citet{cattaneo16_multi}, this approach usually does not fully exploit the multi-cutoff RD structure.  We show that our approach is readily extendable to the case of multiple cutoffs following \citet{cattaneo16_multi} (see, \citet{cattaneo2020stata} and \citet{cattaneo21_multi} for further discussions).  Furthermore,  we show that the multi-cutoff setup reveals further heterogeneity in the empirical application we consider in this paper.  Existing methods for the case of multiple running variables do not examine the case of various cutoffs. To the best of our knowledge, this paper outlines the first actual contribution of an RD framework with multiple running variables and multiple cutoffs.

\begin{figure} 
\caption{SPP bivariate assignment rule}
\centering
\includegraphics[width=.5\textwidth]{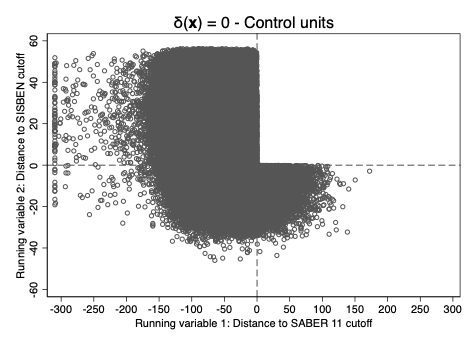}%
\includegraphics[width=.5\textwidth]{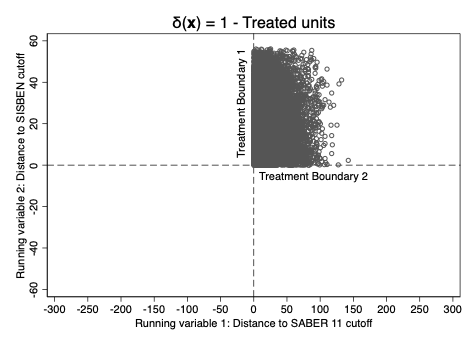}
\caption*{\footnotesize \textbf{Note:} The figure illustrates SPP bivariate assignment rule $\delta(X_{1,i},X_{2,i})=\mathbf{1}\{X_{1,i}>\hat{x}_{1}, X_{2,i}>\hat{x}_{2}\}$. The x-axis contains SABER 11 scores and the y-axis SISBEN scores centered at their respective cutoff breaking units into control (left) and treated observations (right).}
\label{fg:scattspp}
\end{figure}

The bivariate assignment rule for the program takes the for $\delta(X_{1,i},X_{2,i})=\mathbf{1}\{X_{1,i} \geq \hat{x}_{1}, X_{2,i} \geq \hat{x}_{2}\}$ where $X_{1}$ represents SABER 11 scores with its respective cutoff $\hat{x}_{1} = 310$, and $X_{2}$ SISBEN scores with cutoff $\hat{x}_{2} = 57.21$\footnote{ Given that SISBEN's cutoff varies with geographical location (i.e., 57.21 for main cities, 56.32 for other urban areas, and 40.75 for rural areas) each observation is centered at its associated value.}. Figure \ref{fg:scattspp} breaks into two scatter plots the distribution of observations by the assignment rule $\delta({\mathbf{X}})$, where we centered both variables in their respective thresholds. We adopt the following two conventions in the rest of the paper. First, forcing variables are always centered at their cutoff,. Second,  we inverse the scale of the SISBEN score to obtain a poverty index (i.e., higher score higher poverty level) instead of a wealth index.  Both practices simplify the description of results and do not alter the underlying statistical analysis.

The program was announced two months after students took SABER 11 test in August 2014. The announcement date serves as a critical feature of the selection mechanism of the program.  Given that the program was advertised after students took the test, no possible manipulation at the threshold is plausible. Furthermore, the date used for the wealth index was September 2014, where no amends to the household's wealth index were possible. The idea of manipulation at the threshold is especially relevant in RD designs because these kinds of incentive-based schemes might affect the internal validity of the study \citep[see,][]{McCrary08}. 

Ultimately, eligible students had to be admitted to one of the 37 accredited universities in the country, at the time of announcement, which could be either public (20) or private (17), to be eligible for the financial aid.\footnote{ \cite{DNP_2016} and \citet{Londono19} provide further details on the program and policy.}

\subsubsection{Running variables and institutional setting}

The first running variable in this RD design is SABER 11 score.  Although the initial cutoff was 310, the score rose to 318 in the following year (2015) and 348 for the last cohort (2017). There are two reasons for this result. First, new cohorts of students were scoring higher in the test since the cutoff was defined in terms of standard deviation from the mean. The higher cutoff is driven by both an increase in the mean and standard deviation of the distribution. The second is that since the program gained popularity in subsequent years among households and students, every year the take-up ratio increased and more eligible students when applying for financial aid. Consequently, the yearly number of beneficiaries exceeded the target number of 10,000, and the Ministry of Education had to slightly adjust the original number of standard deviations from the mean. To preserve the internal validity of the RD design, we focus the statistical analysis in the next sections on the first cohort of the program (i.e., 2014 test-takers).

The second running variable is SISBEN's wealth index. The instrument is a household targeting system officially used in Colombia to target social programs to the vulnerable and poor \citep{sisben}. SISBEN scores vary between 0 and 100, where lower scores indicate lower socioeconomic conditions of the household. For simplicity on the exposition and interpretation of the results, we inverse the scale to get a poverty index so that both variables are interpreted in a similar fashion where higher scores imply a higher likelihood of being eligible for financial aid. Eligibility cutoffs also vary slightly with geographical location; however, given that we centered both running variables at zero by subtracting the cutoff from each observation, they do not alter the statistical analysis. These cutoffs stayed the same throughout the program.

The final feature of the program is that students must be admitted into at least one of the 37 accredited institutions of the country. These institutions represent around 13 percent of all schools and are certified as high-quality institutions by the National Accreditation Council (CNA for its Spanish acronym). By 2017, and partly explained by the incentive imposed by the program, the number of institutions with high-quality accreditation rose to 44. Among the 44 institutions, 17 were public and 27 private. Colombia's public institutions are highly subsidized and usually entirely free for the lowest-income students. 

We now discuss two additional empirical elements that arise from the institutional setting that play a significant role in the paper's statistical analysis and findings.  On the one hand, since the program offered a forgivable loan upon graduation, students only enjoyed the benefits of a grant if they eventually graduated from college.\footnote{ This requirement is a combination of an imposed legal provision for this kind of program in Colombia, as well as a policy target given a 50 percent dropout rate in the country \citep{MEN_2016}.} The graduation requirement implies that incentives to accept financial aid are not trivial. For instance, financial help became a loan if students did not manage to graduate or drop out of college.  As we will show later in Section \ref{results}, only around 60 percent of all eligible students requested financial assistance in the first year.\footnote{ As previously argued, the low take-up is partly explained by the unexpected introduction of the program as well.} The low take-up ratio induces what is known as imperfect compliance in the research design.  We address this statistical matter by analyzing both sharp and fuzzy multidimensional RD designs that deal with the program's incomplete take-up.

Another feature of the program is that in addition to covering tuition fees, the program also awarded a financial stipend to students. The size of the aid was proportional to the distance from the selected higher education institution to the student's house. For example, students received four monthly minimum wages (MMW) in each academic period\footnote{ Semesters in the case of Colombia.} if they were from a city away from their university,  and only one MMW if they studied in the same metropolitan area where they lived. Distinct values applied for intermediate distances.  Contrary to the graduation requirement,  the stipend would likely incentivize a larger take-up ratio.

These two last features of the program are empirically relevant since they ultimately drive the incentives to join the program and determine the size of the estimated effects. The next section describes the different sources of the data before discussing the empirical results.


\subsection{Data}

We use four main sources of information. The first and primary dataset contains results for the national standardized high school exam SABER 11 for the second semester of 2014. The test is a national requirement for high school graduation, and the information is collected by the ICFES (National Evaluation Agency of Colombia)\footnote{  \textit{Instituto Colombiano para el Fomento de la Educación Superior} for its Spanish name.}. In 2014, around 574,000 students took the test that serves as the first assignment variable in the RD design. The eligibility cutoff for SABER 11 scores is 310/500. Table \ref{tab:sum} presents summary statistics and Figure \ref{fg:sabersisben} (left) in Appendix \ref{sc:app} plots the distribution of the variables with its respective cutoff. The standardized nature of the test is consistent with the almost symmetric distribution of the scores with a mean of 250 and median of 245.

Given that eligibility to the program also requires a score below a specific cutoff in the SISBEN wealth index, data on test taker is matched with that from the SISBEN database for the same year.\footnote{ SISBEN data is collected by the Department of National Planning (DNP), an independent public institution from the ICFES. The autonomy of these organizations imposes an additional challenge when merging both administrative datasets, given that the two sources of information are not fully harmonized. For instance, teenagers in Colombia change their ID number when coming of age (i.e., 18 years old or older), and misspellings of first, middle, and both last names are frequent in these datasets.} Although SISBEN administrative information is not a census-type database, around 35 million individuals out of 47 million Colombians were in the database by the end of 2014. The screening process targets the poor and vulnerable population in the country, and individuals not included in the dataset can request the application of the survey-type instrument to obtain a score \citep[see,][for further details]{sisben}. Table \ref{tab:sum} shows that the SISBEN instrument had screened 363,000 (63 percent) out of the 574,000 students by September 2014. The mean SISBEN score is around 36, with a standard deviation of 18. Figure \ref{fg:sabersisben} (right) in Appendix \ref{sc:app} plots the distribution of the variables with respective cutoffs for main cities, urban, and rural areas. SISBEN and SABER 11 scores together define eligibility into the program and constitute the two running variables in multidimensional RD design.\\

\begin{table}[htbp]
\centering
\caption{\label{tab:sum} Summary statistics}
\resizebox{\textwidth}{!}{\begin{tabular}{l*{1}{cccccccc}}
\toprule \toprule
                    &       N&        Mean&          SD&         Min&         p25&         p50&         p75&         Max\\
\midrule
SABER 11 score            &      574,269&      249.88&       43.16&        0.00&      218.85&      245.38&      276.15&      481.54\\
SISBEN score             &      363,096&       36.28&       18.14&        0.54&       21.91&       34.25&       49.83&       93.76\\
Age (2014)                &      572,107&       17.91&        4.49&        1.00&       16.00&       17.00&       18.00&       63.00\\
Program take-up     &       15,423&        0.59&        0.49&        0.00&        -&        -&        -&        1.00\\
Enrolled (2015-1)   &      574,269&        0.19&        0.39&        0.00&        -&        -&        -&        1.00\\
Female              &      574,269&        0.55&        0.50&        0.00&        -&        -&        -&        1.00\\
Private high school &      549,595&        0.25&        0.44&        0.00&        -&        -&        -&        1.00\\
\bottomrule \bottomrule
\end{tabular}}
\caption*{\textbf{Source:} ICFES (2014), DNP (2014), and MEN (2015).}
\end{table}

There are two additional sources of information used in the analysis.  First, the policy target and main outcome variable is post-secondary enrollment (i.e., access to college or university). To obtain this information, we use additional administrative data on post-secondary education collected by the Ministry of Education. Although there is one official information system for collecting data on higher education in the country known as SNIES\footnote{ SNIES by its Spanish acronym or the National Information System for Higher Education for its equivalent name in English.}, we follow \citet{Londono19} and use a parallel but independent source of information that tracks dropout rates in post-secondary education throughout the country known as SPADIES\footnote{ \textit{Sistema de Prevención y Atención de la Deserción en las Instituciones de Educación Superior} for its Spanish name.}. To preserve consistency with existing research, we use in what follows, and when possible, the data employed and made available with the publication by the \citet{Londono19}. The data is also compatible with the official program's impact evaluation led by the Department of National Planning (DNP) \citep{DNP_2016}.


Finally, given that not all eligible students apply to the program's financial aid, we use administrative information on actual beneficiaries (i.e., compliers) from the ICETEX\footnote{ \textit{Instituto Colombiano de Crédito Educativo y Estudios Técnicos en el Exterior} for its Spanish name.}  and Ministry of Education. ICETEX is the institution in charge of the repayment schemes of the program in case of dropout. It serves as a public financial institution in charge of most higher education scholarships and loan-related resources in the country \citep[see,][]{icetex05}. As illustrated in Table \ref{tab:sum}, in the first year of the program, only around 60 percent of eligible students became SPP beneficiaries.


\section{Results} \label{results}

This section exploits SPP's discontinuous merit- and need-based treatment assignment to identify the program's impact on enrollment. We begin by presenting estimations for the traditional methods that simplify the multidimensional discontinuity by implementing standard RD designs. Furthermore, we offer evidence of the multi-cutoff estimates for SISBEN's scores. 

We then illustrate the ability of the boundary RD methods to exploit the underlying assignment rule to identify potential heterogeneous effects. We compare existing approaches with the flexible alternative proposed in this paper. Estimation reveals discontinuous effects along the treatment boundary that complement standard two-dimensional results.

\subsection{Two-dimensional RD estimates}

The centering approach described by \citet{Wong_etal_2013} is the procedure that seeks to simplify the assignment mechanism the most. Given that this method collapses multiple assignment rules into a single running variable, the RD estimation reduces to a unique point estimate. Table \ref{res1} columns (1) and (2) show the estimated effects following the centering approach for the sharp and fuzzy RD, respectively. The latter internalizes the program's imperfect compliance on the assignment rule by using the program's treatment assignment (i.e., eligibility) as an instrument for the actual treatment status (i.e., take-up). Throughout the analysis and when estimating two-dimensional local linear regression, we follow \citet{Calonico2014} bias-corrected RD estimates with robust variance estimator (second row in Table \ref{res1}) but also present conventional RD effects (first row in Table \ref{res1}).

\begin{figure}[h]
\centering
\caption{Two-dimensional centering approach}
\includegraphics[width=0.59\textwidth]{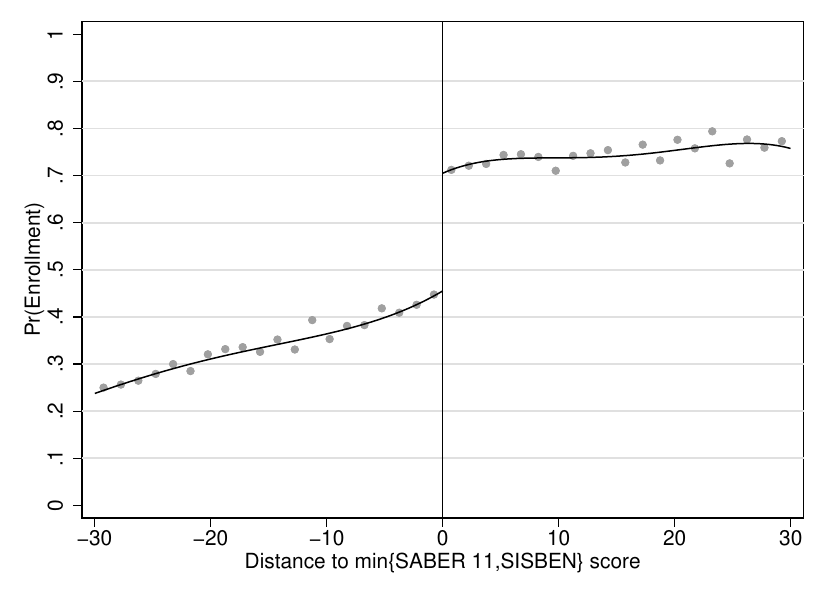}%
\caption*{\footnotesize \textbf{Note:}  The centering approach collapses the two assignment variables (i.e., SABER 11 and SISBEN) into one by taking the minimum value out of the two as the single unit's assignment score. This running variable is plotted on the x-axis. The outcome variable (y-axis) is enrollment in post-secondary education the term after the program's announcement.}
\label{fg:centering}
\end{figure}

Results show that students with scores slightly above zero have an enrollment probability 26 percentage points higher than those somewhat below it.  Figure \ref{fg:centering}, illustrates the standard two-dimensional RD graphical representation. The probability of post-secondary enrollment for the term after the program's announcement was 70 percent for those above the cutoff and around 44 for those just below it. On a basis of 37 percent average enrollment in the next year after the exam, the 24 percentage points gain is equivalent to a 70 percent increase in enrollment. \footnote{ Given that the metric units and scale of the two running variables in our empirical application are different (see, Table  \ref{tab:sum}), we do not suggest following this approach \citep[see,][for further discussions]{Wong_etal_2013}.}

%

\begin{figure}
\caption{Conditional or univariate approach}
\centering
\includegraphics[width=.5\textwidth]{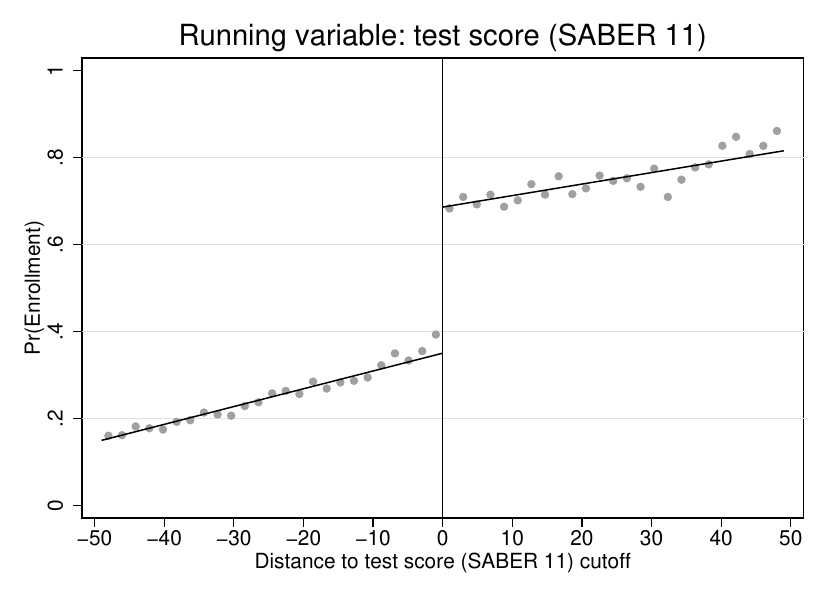}%
\includegraphics[width=.5\textwidth]{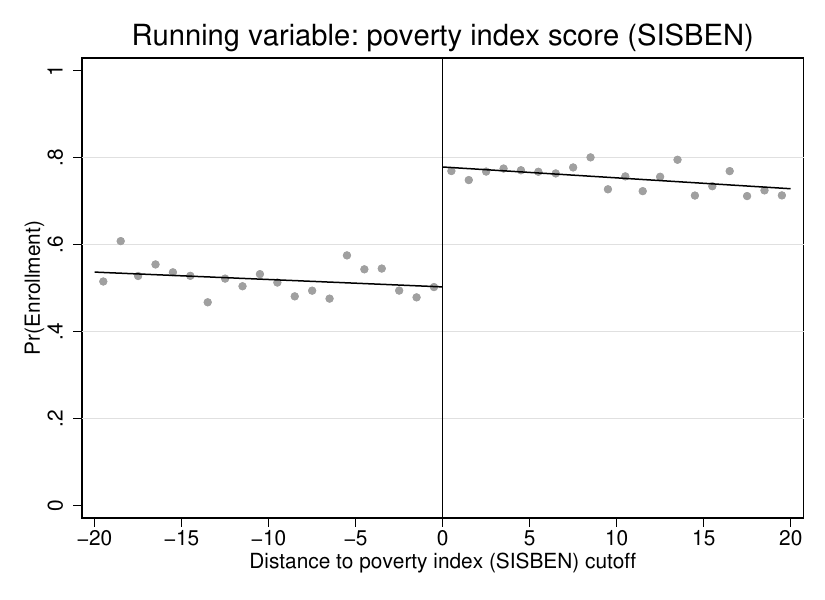}
\caption*{\footnotesize \textbf{Note:} Data-driven RD plots when using SABER 11 (left) and SISBEN (right) as running variable. Dots represent sample average within optimal bins, and lines are global polynomial estimations of order one (1) that approximate the population conditional expectation functions for control (left side of the cutoff) and treated units (right side of the cutoff) independently.}
\label{fg:lon19}
\end{figure}

The next approach in complexity is the conditional or univariate approach. This method evaluates boundary-specific effects and allows the estimation of two independent treatment effects for each running variable instead of a unique effect like the centering procedure. The univariate approach is probably the most widely used in RD designs with two forcing variables. For instance, this is the estimation method followed by \citet{Londono19}, and a previous policy report by \cite{DNP_2016}. The authors implement a standard two-dimensional RD analysis independently for both running variables (i.e., SABER 11 and wealth index scores). In practice, \citet{Londono19} limit the RD estimates to all eligible students by SISBEN (SABER 11) scores when using SABER 11 (SISBEN) as the running variable. Figure \ref{fg:lon19} replicates graphically these results where again a sharp discontinuity shows that students slightly above the threshold, have a higher probability of enrolling in higher education in the next semester after graduation. The effect is visibly higher when using SABER 11 test scores as the running variable (Figure \ref{fg:lon19} - Left) than when using wealth index scores (Figure \ref{fg:lon19} - Right) as we discuss in detail later in the analysis.

The sharp RD estimates in columns (3) of Table \ref{res1} show that for need-eligible students\footnote{ We follow \cite{Londono19} convention and define need-eligible as those eligible by wealth index scores and merit-eligible as those eligible by SABER 11 scores.}, a test score slightly above the cutoff raises enrollment by 31.9 percentage points for the bias-corrected RD estimator or 32.3 for the conventional one.\footnote{ The conventional estimate is equivalent to the none bias-corrected RD estimates with conventional variance \citep[see,][]{Calonico2014}.} For merit-eligible students, results in columns (5) of Table \ref{res1} show that wealth index scores above the cutoff raise enrollment by 27.4 percentage points for the bias-corrected or 26.9 for the conventional estimator.

\begin{table}
\centering
\small
\def\sym#1{\ifmmode^{#1}\else\(^{#1}\)\fi}
\caption{Two-dimensional RD estimates for enrollment in post-secondary education \label{res1}}
\resizebox{\textwidth}{!}{\begin{tabular}{l*{6}{D{.}{.}{-1}}}
\toprule
            & \multicolumn{2}{c}{Centering} &\multicolumn{2}{c}{SABER 11} &\multicolumn{2}{c}{SISBEN} \\ 
            &  \multicolumn{2}{c}{All} &\multicolumn{2}{c}{Need-eligible} &\multicolumn{2}{c}{Merit-eligible} \\ 
            &\multicolumn{1}{c}{(1)}&\multicolumn{1}{c}{(2)}&\multicolumn{1}{c}{(3)}&\multicolumn{1}{c}{(4)}&\multicolumn{1}{c}{(5)}&\multicolumn{1}{c}{(6)}\\            
            &\multicolumn{1}{c}{Sharp}&\multicolumn{1}{c}{Fuzzy}&\multicolumn{1}{c}{Sharp}&\multicolumn{1}{c}{Fuzzy}&\multicolumn{1}{c}{Sharp}&\multicolumn{1}{c}{Fuzzy}\\
\midrule
\addlinespace
Conventional&    0.265\sym{***}   & 0.465\sym{***}  & 0.323\sym{***}& 0.587\sym{***} &       0.269\sym{***}&     0.434\sym{***}             \\
            & (0.013) &  (0.021) &     (0.010)         &         (0.016)            &      (0.023)         &         (0.034)            \\
\addlinespace
Bias-corrected     &    0.260\sym{***}   & 0.458\sym{***}  & 0.319\sym{***}& 0.577\sym{***}                      &       0.274\sym{***}& 0.445\sym{***}                     \\
            &   (0.015)  & (0.025) & (0.011)         &         (0.018)            &      (0.027)         &                     (0.040) \\
\addlinespace
\midrule
\multicolumn{1}{l}{\(N\)} & \multicolumn{1}{r}{363,096} & \multicolumn{1}{r}{363,096} & \multicolumn{1}{r}{299,475}&\multicolumn{1}{r}{299,475}&\multicolumn{1}{r}{23,132}&\multicolumn{1}{r}{23,132}\\
\bottomrule
\multicolumn{5}{l}{\footnotesize Standard errors in parentheses}\\
\multicolumn{5}{l}{\footnotesize \sym{*} \(p<0.05\), \sym{**} \(p<0.01\), \sym{***} \(p<0.001\)}\\
\end{tabular}}
\end{table}

Similarly, Table \ref{res1} columns (4) and (6) present the corresponding results for the fuzzy RD estimates. Effects are around 57 percentage points increase when the imperfect compliance is internalized, and test scores (SABER 11) are used as the running variable and approximately 44 when the poverty index serves as the forcing score.  \citet{Londono19} shows that such an enrollment impact virtually eliminates the socioeconomic status (SES) enrollment gradient among top decile test-takers in Colombia.

We can further extend the conditional approach to exploit the multi-cutoff structure of the poverty index. In Appendix \ref{sc:app2}, we estimate a simple multi-cutoff RD where we allow for independent effects by location. We show that the program's impact on students that faced the threshold in rural areas is not statistically different from zero.  Figure \ref{fg:multicutoffs} in Appendix \ref{sc:app2}, shows the equivalent graphical discontinuity when the different cutoffs are not pooled into one single score like in Figure \ref{fg:lon19} (left).

In what follows, we exploit the multidimensional assignment rule to identify a more comprehensive set of treatment effects along the entire boundary. We show results for the flexible boundary approach proposed in Section \ref{mer} and compare it with existing methods.

\subsection{Boundary RD estimates}

\subsubsection{Semiparametric approach by \citet{Papay2011}}

\begin{figure} 
\centering
\caption{Semiparametric approach by \citet{Papay2011}}
\includegraphics[width=.5\textwidth]{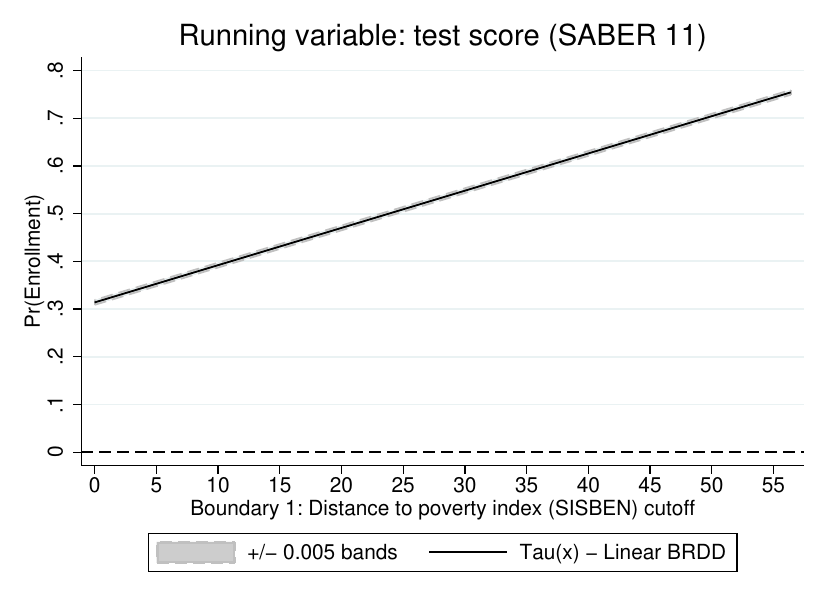}%
\includegraphics[width=.5\textwidth]{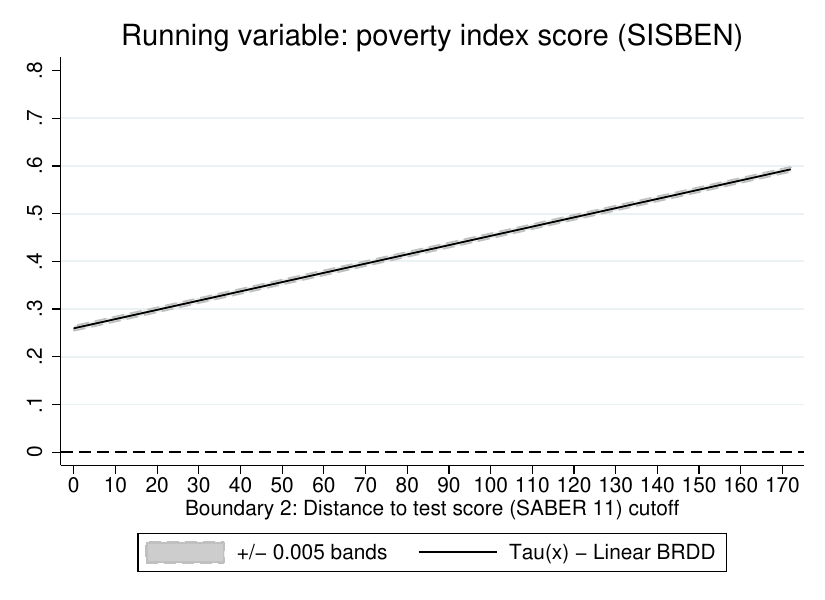}
\caption*{\footnotesize \textbf{Note:} Following the specification proposed by \cite{Papay2011}, the figure shows linear estimation for both SABER 11 (left) and SISBEN (right) forcing variables. }
\label{fg:pwm11}
\end{figure}

Following the semiparametric approach by \citet{Papay2011}, Figure \ref{fg:pwm11} presents the estimations for both treatment boundary. As in our previous analysis, the y-axis still shows the policy outcome of interest $\text{Pr}(\text{Eronllment})$ which equals the probability of post-secondary enrollment. However, the x-axis plots the discontinuous jump along each treatment boundary ($\mathbb{B}_1$ and $\mathbb{B}_2$), as previously illustrated in Figure \ref{fg:brdd} (in three-dimensions) and \ref{fg:scattspp} (in two-dimensions). In the case of boundary $\mathbb{B}_1$ (boundary $\mathbb{B}_2$), the variable that induces a discontinuous jump along the edge is the test score (poverty index score), and all units represent need-eligible (merit-eligible) students with corresponding distances to the policy cutoff (see, Figure \ref{fg:pwm11}).

Results show that scores slightly above the cutoff raise enrollment by 31 percentage points on boundary 1 (Figure \ref{fg:pwm11}, left) and 26 on boundary 2 (Figure \ref{fg:pwm11}, right). These estimates are consistent with the results for the univariate approach. However, the estimations seem inconsistent when moving away from the eligibility cutoffs. For instance, the further away from the SABER 11 cutoff (i.e., for higher SABER 11 scores), the larger the implied impact of the program by the parametric linear estimation (Figure \ref{fg:pwm11}, left). With the largest impact of almost 75 percentage points higher probability of enrollment in post-secondary education. A similar pattern is estimated for SISBEN scores away from the threshold (i.e., for higher poverty indexes) with the largest effect estimated at around 60 percentage points higher probability of enrollment (Figure \ref{fg:pwm11}, right).

The estimations show one of the main limitations of the semiparametric approach. The linearity imposed by the parametric specification implies that effects move monotonically away from the threshold.\footnote{ Unless the slope is 0, where the effect is constant, and no need for boundary RD estimations are needed.} This characteristic is not always desirable since heterogeneous effects along the boundaries might be nonlinear as we show for the nonparametric approaches in the next sections.\footnote{ Even though we show that these effects are mainly incompatible with the nonparametric methods, the results nevertheless highlight the ability of boundary RD designs to estimate a broader set of treatment parameters.} We do not discuss confidence interval estimator or inference procedures for treatment effects along the boundary because they are not developed in the original paper \citep[see,][]{Papay2011}.

A final remark is that the semiparametric approach is quite sensitive to the bandwidth selection. For instance, Figure \ref{fg:pwm11_2} in Appendix \ref{sc:app} replicates the results for a bandwidth 25 percent higher than the one in Figure \ref{fg:pwm11}. The results show that the slope for both boundaries becomes negative and almost flat for the boundary along SISBEN scores. We do not recommend to follow the semi parametric approach by \citet{Papay2011} when nonlinear effects are expected. For consistency with the nonparametric methods that we study in the next sections, we use for the results in Figure \ref{fg:pwm11} the mean optimal bandwidth selected by the nonparametric approach describe in Section \ref{mer}.

\subsubsection{Alternative nonparametric approach}

\begin{figure}
\centering
\caption{Alternative nonparametric approach - Boundary RDD plot}
\includegraphics[width=.5\textwidth]{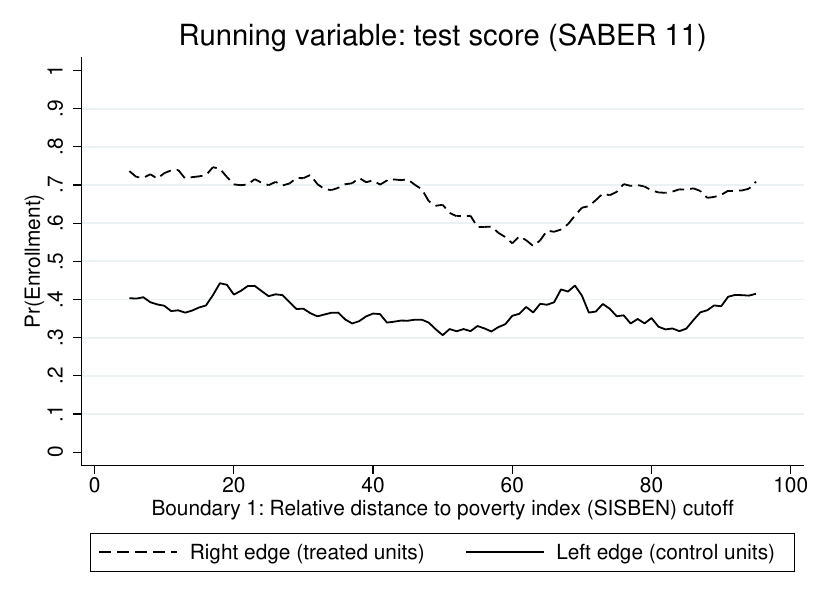}%
\includegraphics[width=.5\textwidth]{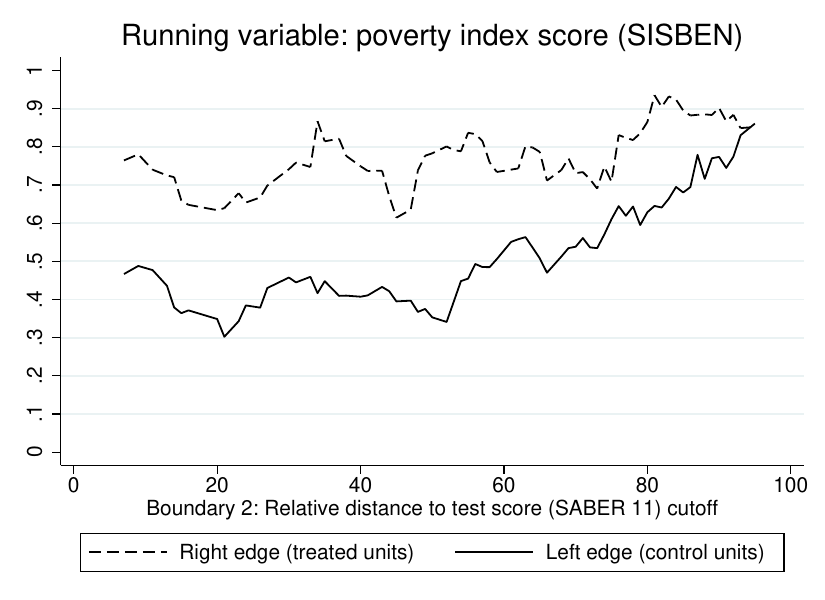}
\caption*{\footnotesize \textbf{Note:} Estimated edges following the flexible nonparametric approach described in Section \ref{mer} for a moving window that takes 10 percent of the observations at each point along the treatment boundary. The bold line plots the bias-corrected local linear regression for control units (i.e., left side of the boundary) and the dotted line the bias-corrected local linear regression for treated units (i.e., right side of the boundary). }
\label{fg:mer1}
\end{figure}

\begin{figure}
\centering
\caption{Alternative nonparametric approach - Sharp BRDD $\tau_{SBRD \lvert \mathbb{B}_j}(\mathbf{x})$}
\includegraphics[width=.5\textwidth]{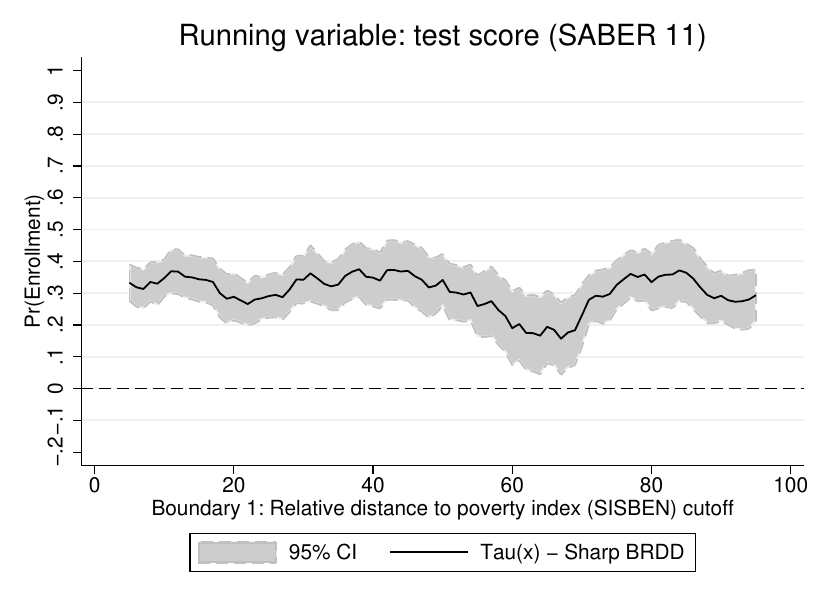}%
\includegraphics[width=.5\textwidth]{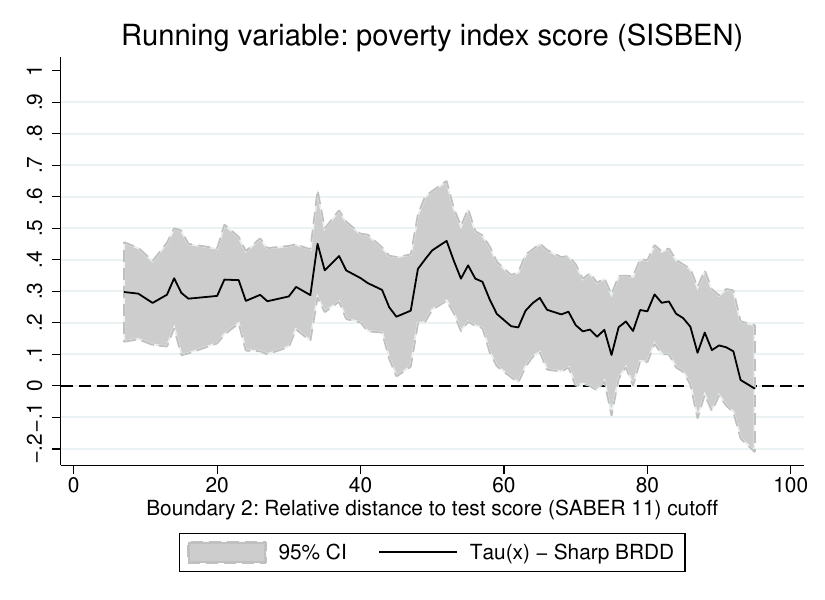}
\caption*{\footnotesize \textbf{Note:} Boundary RD estimates following the flexible nonparametric approach described in Section \ref{mer} and equation (\ref{eq:sbrd_new}) for a moving window that takes 10 percent of the observations at each point along the treatment boundary. The black line plots the bias-corrected local linear regression estimate and the gray area the 95 percent bias-corrected robust confidence interval.}
\label{fg:mer19}
\end{figure}

To illustrate the boundary RD benefits, we first show how the bias-corrected local linear regressions estimate the probability of post-secondary enrollment for control and treated units along the two boundaries.  In all the plots for boundary estimations that follow, we use relative distance to the cutoff using the percentile distribution of each running variables. Figure \ref{fg:mer1} (left) shows that although the likelihood of enrollment is mainly homogenous for both control (around 70 percent) and treated units (about 40 percent), some subpopulations are affected differently by the policy. For instance, students with relative distance to poverty index around 60, face treatment effects (i.e., vertical discontinuity) that are smaller (around 15 percentage points) than for those with scores close to the cutoff (about 30 percentage points).  Figure \ref{fg:mer1} displays the multidimensional equivalent to the standard RD plots. Nevertheless, in the three-dimensional case, there is not a single jump in the outcome of interest but rather a continuous discontinuity along the boundary.  Overall, we can conclude that the vertical distances for boundary $\mathbb{B}_1$ are rather homogenous throughout the frontier (see,  Figure \ref{fg:mer1} (left) and Figure \ref{fg:mer19} (left)).

On the other hand, Figure \ref{fg:mer1} (right) shows that for the second treatment boundary, the probability of enrollment is somewhat more heterogeneous. For example, although the likelihood of enrollment in post-secondary education is around 76 percent (46 percent) for treated units (control units) close to the test score cutoff, this probability goes up to around 85 percent for treatment and control unites far away from the cutoff (i.e., students with the highest test scores).  Strikingly, the two probabilities converge to the same values, suggesting no treatment effect for the top 1 percent of the score distribution. For the results in Figure \ref{fg:mer1} and \ref{fg:mer19}, we estimate the percentiles for the test scores distribution above the threshold using all observations, including students without SISBEN scores. 

Figure \ref{fg:mer19} shows the corresponding treatment effects (i.e., vertical discontinuity) following equation (\ref{eq:sbrd_new}) for the two boundaries with a 95 percent confidence interval. Unlike the parametric approach, it is worth mentioning that the estimated effects are moderately nonlinear along the treatment boundary. The ability to capture this heterogeneity is what motivated the study on boundary estimations in this paper.

Consistent with the univariate approach, scores slightly above the threshold raise enrollment by around 33 percentage points in the case of boundary 1 (Figure \ref{fg:mer19}, left), and 28 percentage points for boundary 2 (Figure \ref{fg:mer19}, right). Nevertheless, the nonparametric approach also reveals additional heterogeneous effects throughout the frontier compared to the two-dimensional RD estimates and the semiparametric procedure discussed so far. For instance, unlike the estimation for boundary 1 (Figure \ref{fg:mer19}, left) where effects are highly homogeneous and more precisely estimated with narrower confidence interval bands, the impacts for boundary 2 (Figure \ref{fg:mer19}, right) are somewhat more heterogeneous and even non-significant for students with the highest test scores.

Although the effects in Figure \ref{fg:mer19} (left) are consistent with a average treatment effect of approximately 30 percent, students in the upper-middle of the distribution face a treatment effect as low as 15 percentage points increase in the probability of post-secondary education. This effect is equivalent to half the size estimated by the univariate approach. 

Likewise, Figure \ref{fg:mer19} (right) also verify the results observed for the two-dimensional RD effects (see, Table \ref{res1} and Figure \ref{fg:lon19}).  The effect is, on average, smaller and noisier along the entire boundary. \footnote{ \cite{Londono19} also argue that the results when using SISBEN as the running variable are indeed smaller and noisier for the univariate case.} The fact that the estimated effects become statistically equal to zero for students with scores in the upper part of the distribution suggests that what we observe in the univariate RD estimates might sometimes be a combination of significant and non-significant effects along the treatment boundary. This result is also compatible with the admission system in post-secondary education where students with the highest scores are more likely to find alternative financial aids (i.e., direct scholarships with universities) that make financial assistance in the form of a loan less advantageous for them.


Figure \ref{fg:mer19} also shows one potential limitation of the approach. The limited sample size for local estimations along the boundary might induce treatment effects to exhibit some local noise or jumps relative to methods that impose some additional smoothing \citep[see,][for an example]{Zajonc2012}. One way to minimize local disturbance is by increasing the estimation window ($F_{\eta}$) and hence adding more data points for each nonparametric estimation.\footnote{ This could be somewhat problematic for estimating effects at the tails of the distributions. } Figures \ref{fg:mer1_20}, \ref{fg:mer19_20},  and \ref{fg:mer19fuzzy_20} replicates the results of Figure \ref{fg:mer1}, \ref{fg:mer19}, and \ref{fg:mer19fuzzy} by using a 20 instead of a 10 percent window.  Results are indeed smoother,  but they reduce the prospect of investigating treatment effects at the end of the distribution. Although we do not consider them problematic in this empirical application, research might consider some additional smoothing in other settings. Mean treatment effects for each boundary (30 percentage points for Figure \ref{fg:mer19}  (left) and 27 percentage points for Figure \ref{fg:mer19} (right)) are, as expected, consistent with the univariate approach follow by \citet{Londono19}.

Figure \ref{fg:mer19fuzzy} in Appendix \ref{sc:app} shows the equivalent fuzzy boundary RD estimates following our approach in equation (\ref{eq:fbrd_new}). The results are essentially consistent with the sharp estimates; however, treatment effects are scaled by the imperfect complies in an IV-fashion. Given that results are mostly aligned with the ones in Figure \ref{fg:mer19}, and that the denominator in equation (\ref{eq:fbrd_new}) seems to follow a similar pattern as the sharp discontinuity, we continue the analysis following the results for the sharp boundary RD.

As a robustness check,  Figures \ref{fg:mer1_5} and \ref{fg:mer19_5} in Appendix \ref{sc:app} replicates the analysis presented in this section but for a smaller window of 5 percent of the data around each point along the boundary. The bias-corrected RD estimates are partially consistent with Figures \ref{fg:mer1} and \ref{fg:mer19} with two main differences. First,  Figure \ref{fg:mer19_5} (right) shows that for the second boundary, the estimated effects are rather noisy with larger confidence intervals.  This is expected for two reasons.  There is a reduction in sample at each point to perform the local linear regressions and we also know from previous results for the two-dimensional estimates that the effects for this boundary (i.e., poverty index) are relatively noisier \citep[see,][]{Londono19}. Second, Figure \ref{fg:mer19_5} (left) shows that although results are more uniform for this boundary,  the 15 percentage points increase in the probability of enrollment for those in the upper-middle part of the distribution becomes somewhat non-significant from zero\footnote{ This result is similar to the one we find in Section \ref{Zajonc2012} following \citet{Zajonc2012} method.}. Even though we consider these estimates rather noisy and decide to stick to our preferred specifications for a 10 percent window, similar results might suggest the likelihood of heterogeneous effects even when boundary effects are largely homogenous and precisely estimated at different boundary points, as is the case when using the test score (SABER 11) as running variable.

Overall,  we have shown that by flexibly exploiting the multidimensional assignment rule,  policymakers and researchers can learn about the decomposition of treatments effects along the two dimensions of the assignment mechanism.

\subsubsection{Nonparametric method by \citet{Zajonc2012}} \label{Zajonc2012}

\begin{figure} 
\centering
\caption{The nonparametric approach by \citet{Zajonc2012}}
\includegraphics[width=.5\textwidth]{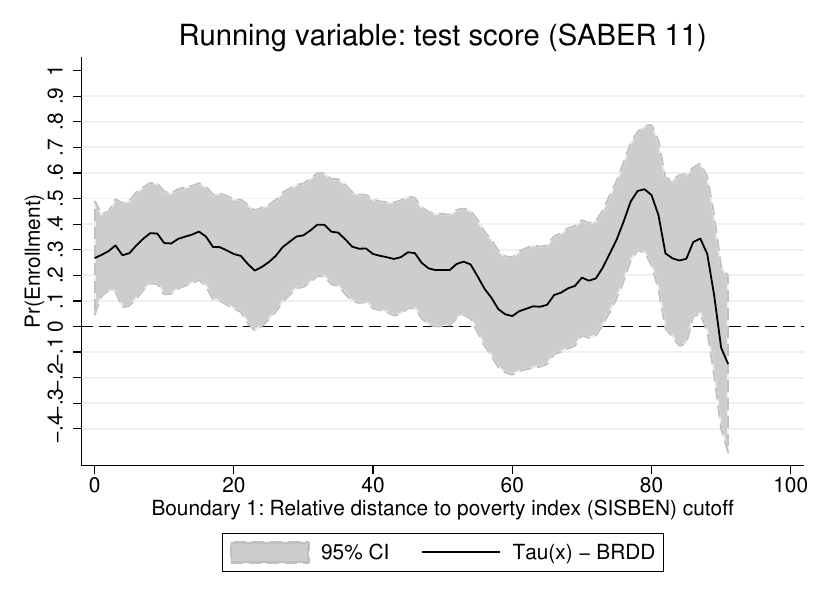}%
\includegraphics[width=.5\textwidth]{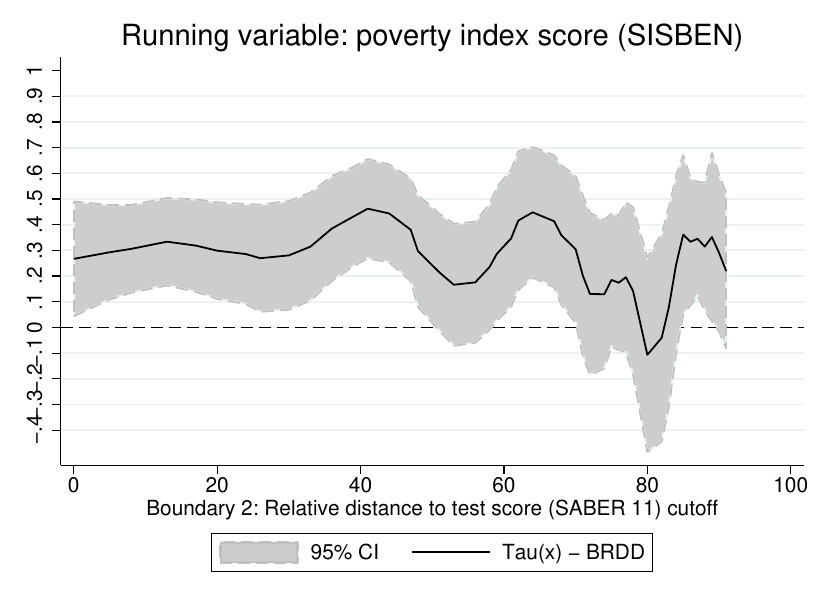}
\caption*{\footnotesize \textbf{Note:} The black line plots the bivariate local linear regression estimates and the gray area the 95 percent robust confidence interval. The optimal rule-of-thumb bandwidth ($h^*_{ROT}$) is selected as the average optimal bandwidth $\hat{h}_{\text{opt}}(\mathbf{x})$ estimated for an evenly spaced grid along the boundary.  We drop treatment effects for the top 9 percent of the distributions, as point estimates and confidence intervals become noisy and do not fit the plot correctly. }
\label{fg:zaj12}
\end{figure}

Figure \ref{fg:zaj12} shows results for the nonparametric method by \citet{Zajonc2012}. Given that the estimated optimal data-dependent bandwidth selected by Zajonc's procedure seems too small and estimates are highly noisy (see, Figure \ref{fg:zaj12_min} in Appendix \ref{sc:app} for the original results), we decided to present and discuss results for a slightly modified version of their bandwidth. The difference is that instead of selecting as the optimal rule-of-thumb bandwidth ($h^*_{ROT}$) the minimum optimal bandwidth $\hat{h}_{\text{opt}}(\mathbf{x})$ estimated for an evenly spaced grid along the boundary, we select an optimal bandwidth as the mean optimal bandwidth out of all the estimated $\hat{h}_{\text{opt}}(\mathbf{x})$. With this modification, results become smoother and consistent with our alternative nonparametric procedure addressed above.

Similar to our nonparametric approach, the estimated effects in Figure \ref{fg:zaj12} show some heterogeneity no just within each boundary but also between them ($\mathbb{B}_1$ and $\mathbb{B}_2$). Two main differences emerge. First, the estimated effects in Figure \ref{fg:zaj12} (left) get a bit noisier than the ones for the alternative flexible approach (Figure \ref{fg:mer19}, left). Second, the point estimates for the second boundary (Figure \ref{fg:zaj12}, right) are smoother for observation closer to the cutoff but less precisely estimated for the upper part of the distribution.\footnote{ Figure \ref{fg:zaj12} and Figure \ref{fg:zaj12_min} drop treatment effects for the top part of the distribution,  as point estimates and confidence intervals become noisy and do not fit the plot appropriately. } Mean treatment effects for each boundary (25 percent in Figure \ref{fg:zaj12} (left) and 27 percent Figure \ref{fg:zaj12} (right)) are slightly smaller than the univariate and alternative nonparametric approach.

It is worth noticing three features of Zajonc's method that partially explain the differences in the results. First, \citet{Zajonc2012} follows a fixed bandwidth for both boundaries, which is selected as the smallest plug-in bandwidth from an evenly spaced grid of points. Second, the weighting scheme used in the bivariate local linear estimations that weight both running variables at the same time \citep[see,][ for discussions]{Zajonc2012}. In our flexible nonparametric approach, the bandwidth is flexibly estimated along the boundary, and the weighting scheme only involves one of the forcing variables at the same time (i.e., the variable that induces a discontinuous jump). Finally, our inference approach uses bias-corrected RD estimates with robust variance, and \citet{Zajonc2012} follows  RD estimates with conventional robust variance.  Overall, the two procedures capture similar patterns along the treatment boundaries.


\section{Conclusions}

In this paper, we study an extended regression-discontinuity (RD) design where assignment rules involve more than one running variable at the same time. We review the existing literature on multidimensional RD designs and propose an alternative flexible nonparametric approach to estimate the multidimensional discontinuity by univariate local linear regression.

We show that if treatment assignment involves more than one covariate simultaneously (i.e., a vector of covariates with different cutoffs), traditional RD estimates might not fully account for heterogeneous treatment effects along the treatment boundary. In particular, RD design with multiple assignment variables identifies conditional effects at every point along the treatment boundary rather than at a single point \citep{Zajonc2012}. We evaluate the performance of the boundary RD design with an empirical application to a merit- and need-based scholarship program for low socioeconomic students in Colombia. The program's key feature is that it selects students into the program using a merit-based (i.e.,  a score above a cutoff in the national standardized exam) and need-based (i.e., wealth index below a cutoff) assignment rule.

Our results show that exploiting the multidimensional nature of the treatment assignment for the program reveals partial heterogeneous effects along the treatment boundaries. Compared to standard two-dimensional RD estimates, our nonparametric approach shows that estimated effects could be 50 percent smaller for some sub-population exposed to the policy when studying the need-based treatment boundary. Furthermore, two-dimensional RD estimates could also involve endogenous averaging of significant and non-significant effects for different populations affected by the program when analyzing the merit-based treatment boundary. 

Overall,  by flexibly exploiting multidimensional assignment rules, boundary RD designs could complement standard methods and help policymakers and researchers learn about the decomposition of treatment effects implied by the assignment mechanism.

\bibliographystyle{chicago}
\bibliography{Merlano_2023_WP_BRDD.bib}

\newpage

\appendix
\renewcommand\thefigure{\thesection.\arabic{figure}}    

\section{Supplementary Material}\label{sc:app}
\setcounter{figure}{0}    
%
\begin{figure}[h]
\centering
\caption{SABER 11 and SISBEN distributions}
\includegraphics[width=.5\textwidth]{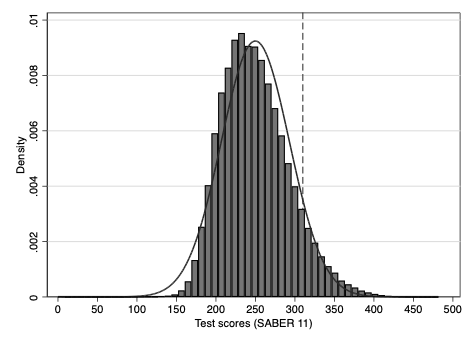}%
\includegraphics[width=.5\textwidth]{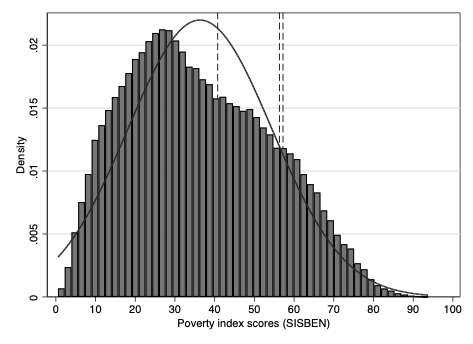}
\caption*{\footnotesize \textbf{Note:}  Figure \ref{fg:sabersisben} (left) plots test scores (SABER 11) scaled to density units for all 2014-II test-takers (N=574,269) with its cutoff point at 310. Figure \ref{fg:sabersisben} (left) plots poverty index scores (SISBEN) scaled to density units for the students screened by the SISBEN instrument (N= 363,096) with its cutoff points. Eligibility cutoffs for SISBEN varied with geographical location with scores below 57.21 for students in the 14 metropolitan areas (i.e., main cities in the country), below 56.32 for other urban areas, and below 40.75 for rural areas. Both plots include an appropriately scaled normal density (solid lines).}
\label{fg:sabersisben}
\end{figure}

%

\begin{figure}[h]
\centering
\caption{The parametric approach by \citet{Papay2011}}
\includegraphics[width=.5\textwidth]{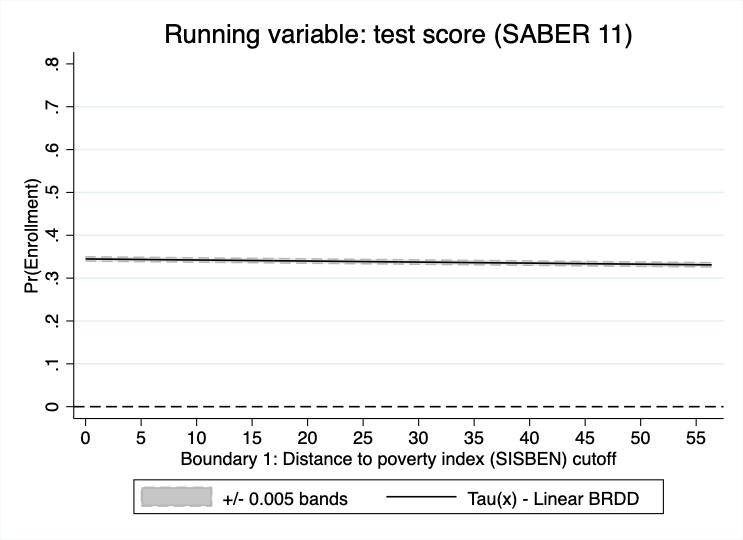}%
\includegraphics[width=.5\textwidth]{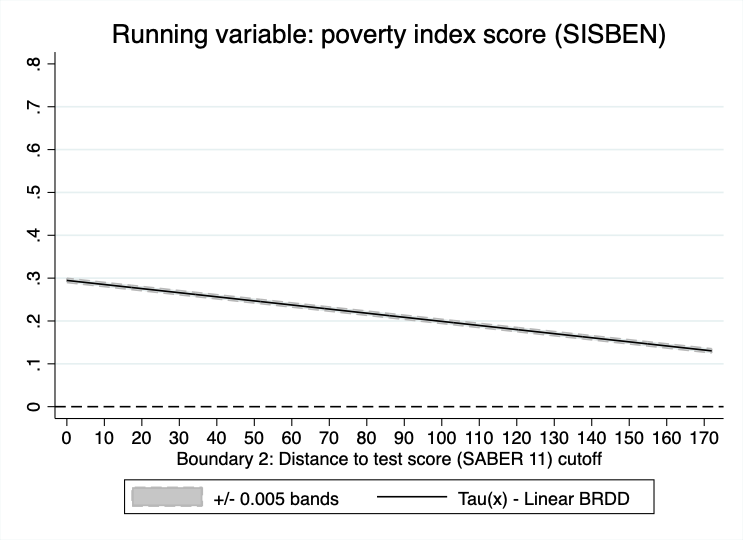}
\caption*{\footnotesize \textbf{Note:} Following the specification proposed by \cite{Papay2011}, the figure shows linear estimation for both SABER 11 (left) and SISBEN (right) forcing variables. }
\label{fg:pwm11_2}
\end{figure}

%

\begin{figure}
\centering
\caption{Alternative nonparametric approach - Fuzzy BRDD $\tau_{FBRD \lvert \mathbb{B}_j}(\mathbf{x})$}
\includegraphics[width=.5\textwidth]{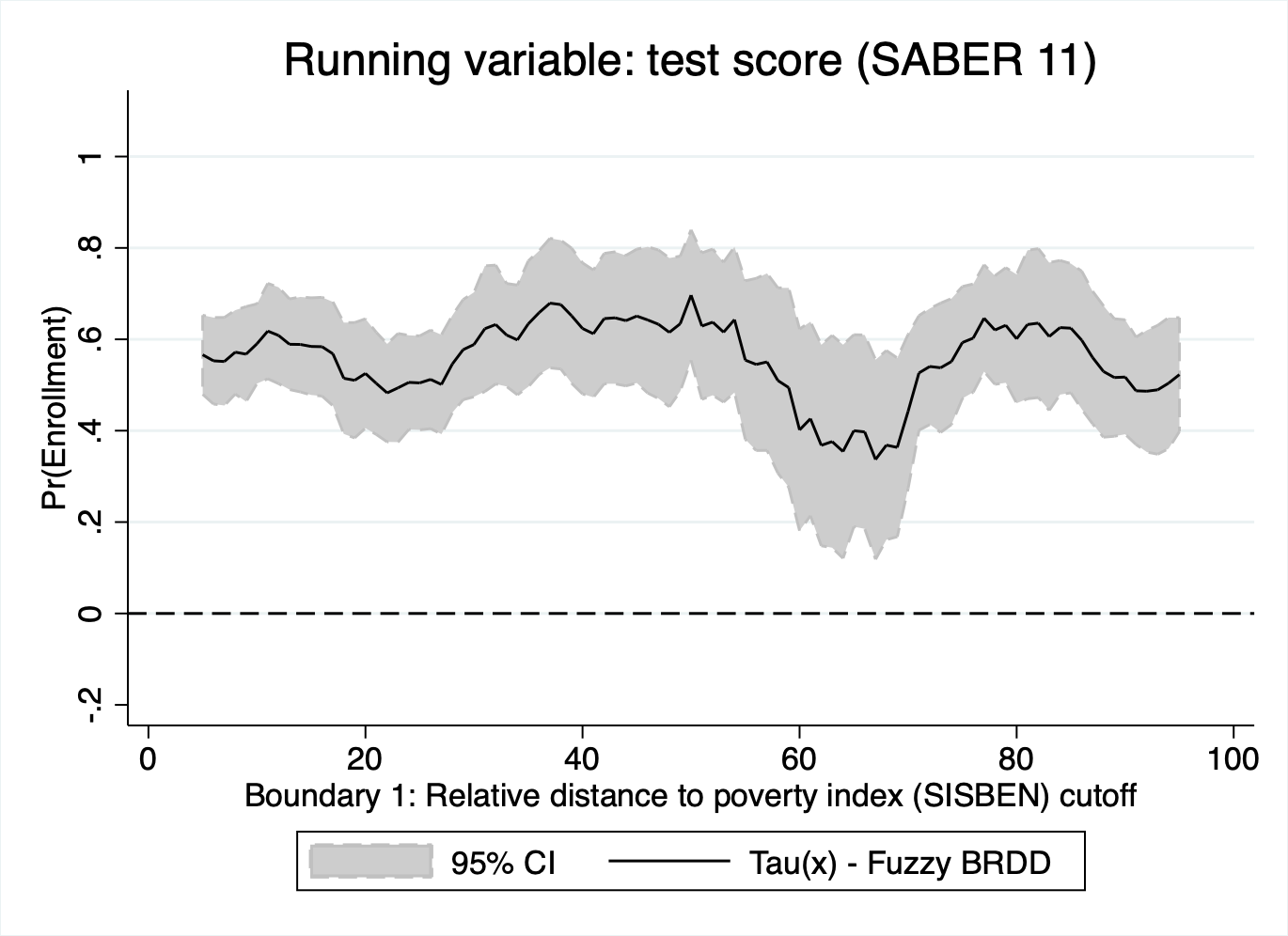}%
\includegraphics[width=.5\textwidth]{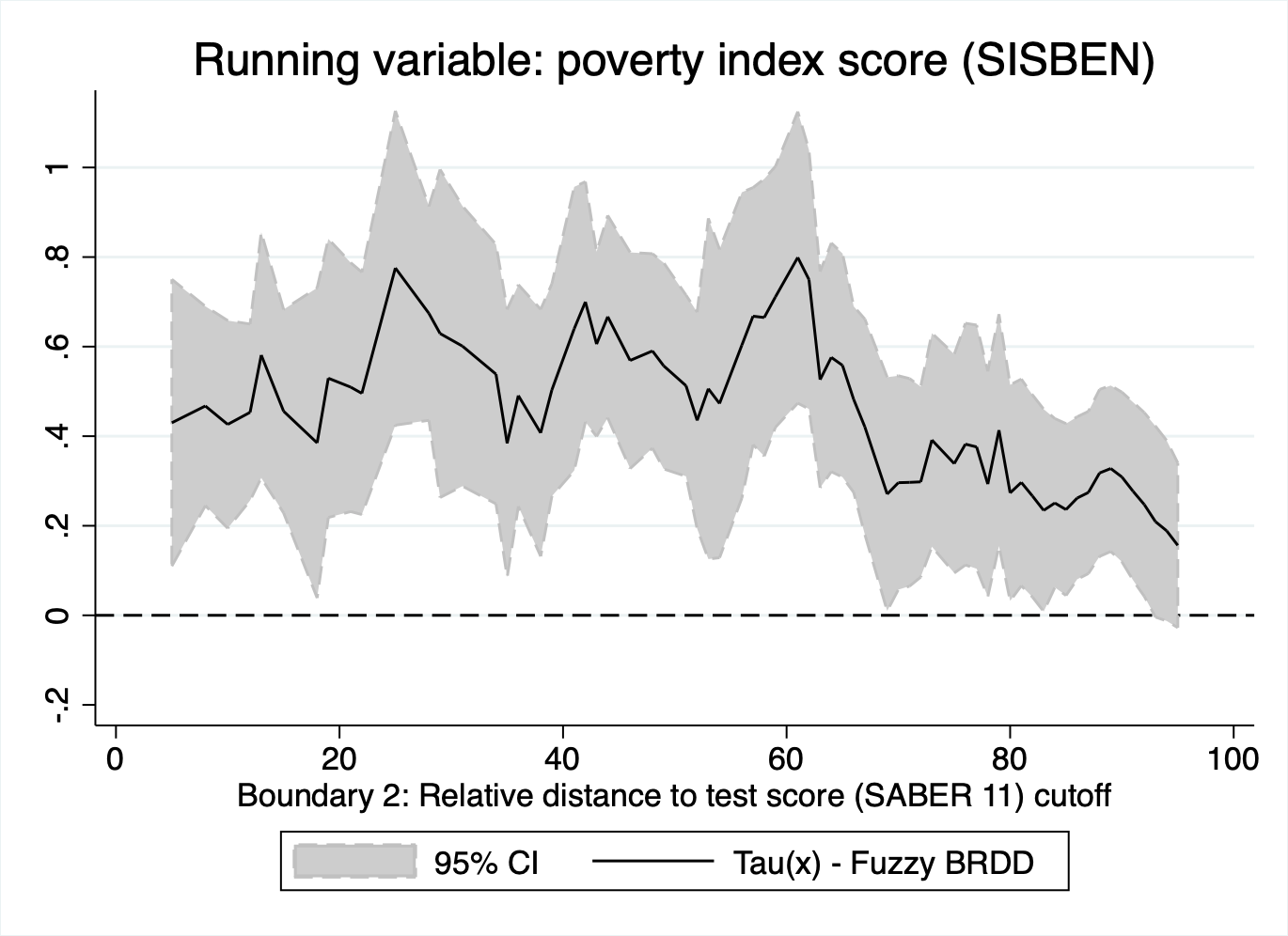}
\caption*{\footnotesize \textbf{Note:} Fuzzy boundary RD estimates following the flexible nonparametric approach described in Section \ref{mer} and equation (\ref{eq:fbrd_new}) for a moving window that takes 10 percent of the observations at each point along the treatment boundary. The black line plots the bias-corrected local linear regression estimate and the gray area the 95 percent bias-corrected robust confidence interval.}
\label{fg:mer19fuzzy}
\end{figure}

\begin{figure}
\centering
\caption{Alternative nonparametric approach - Boundary RDD plot (20\% window)}
\includegraphics[width=.5\textwidth]{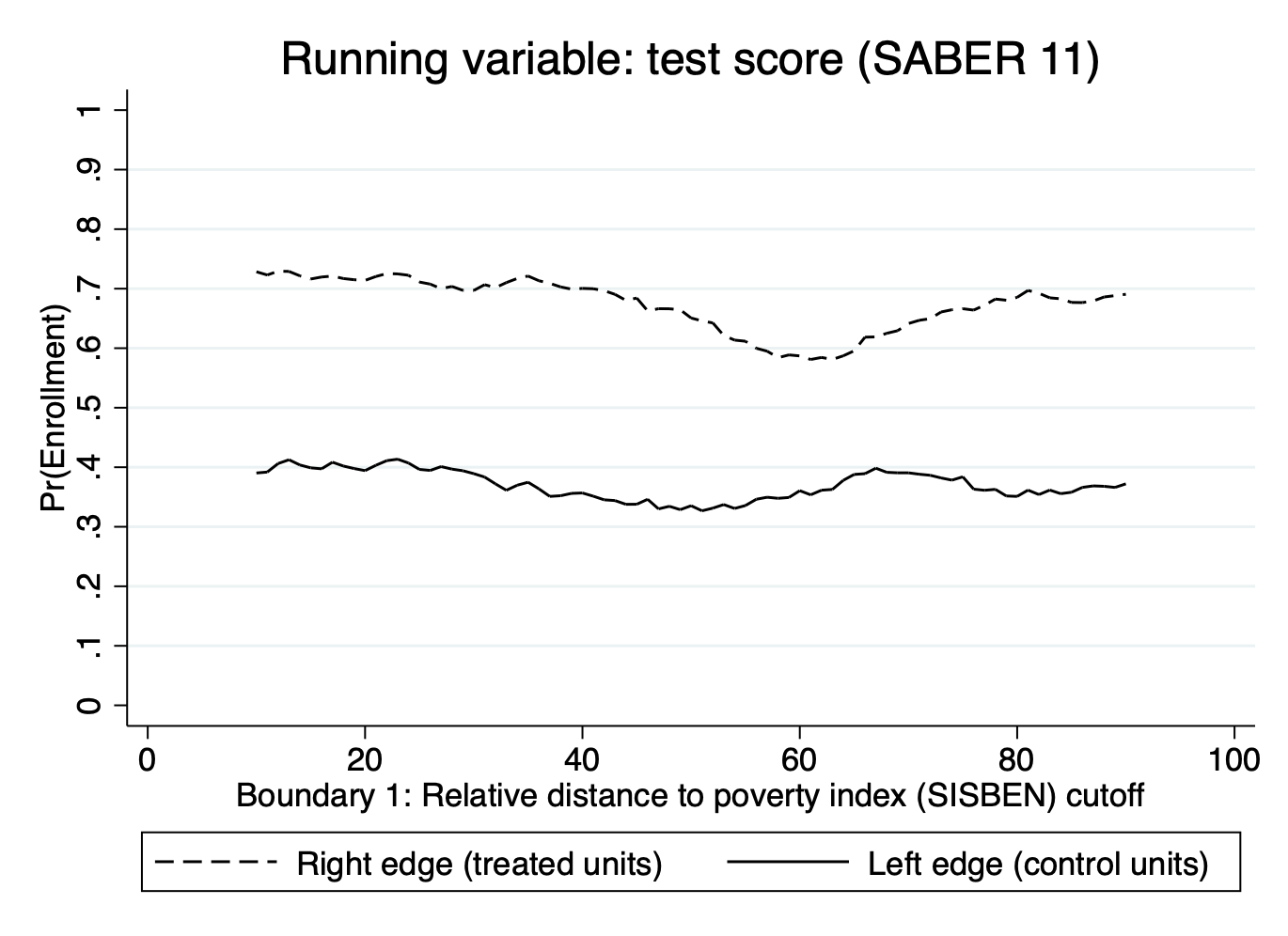}%
\includegraphics[width=.5\textwidth]{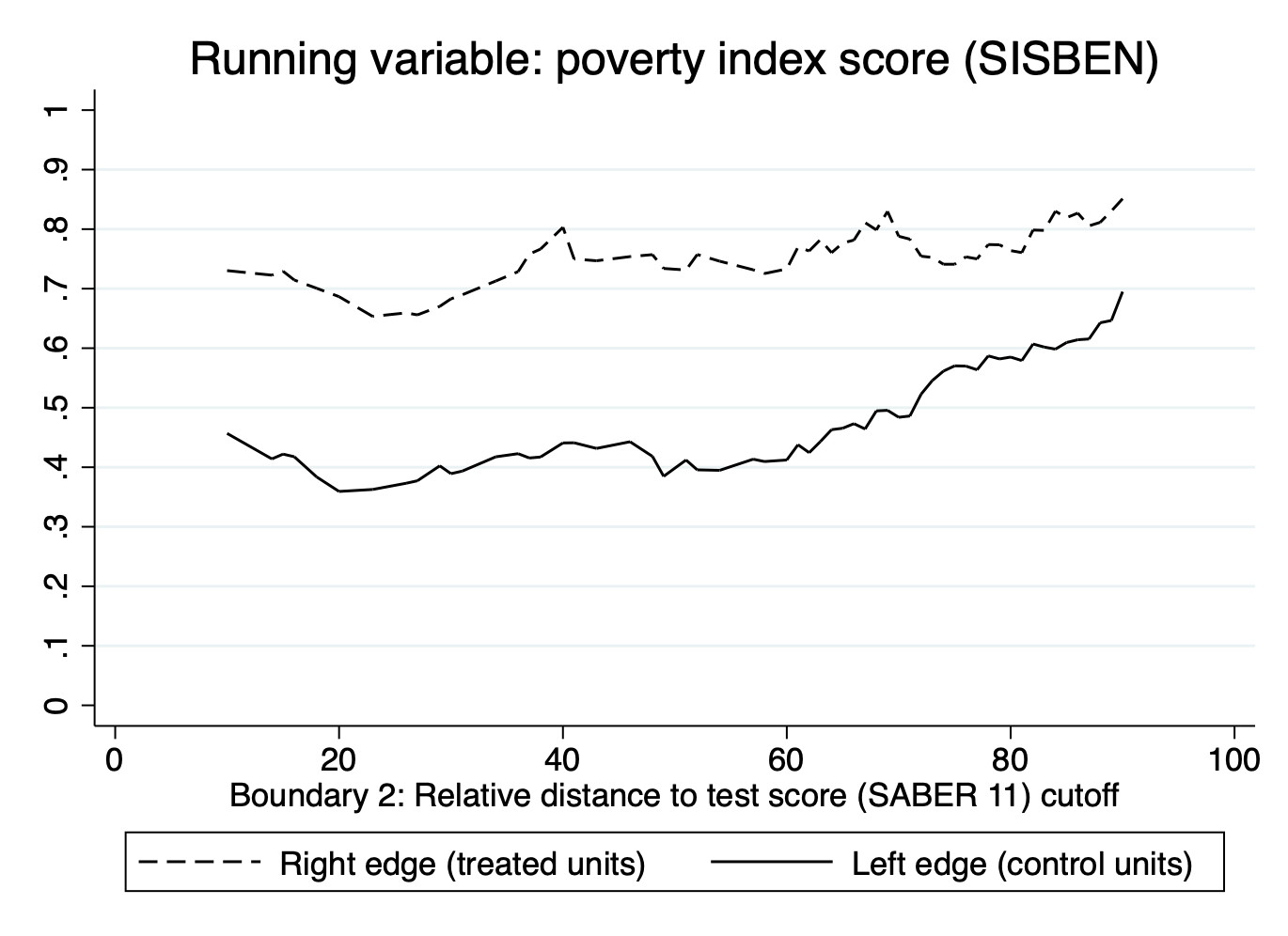}
\caption*{\footnotesize \textbf{Note:} Estimated edges following the flexible nonparametric approach described in Section \ref{mer} for a moving window that takes 20 percent of the observations at each point along the treatment boundary. The bold line plots the bias-corrected local linear regression for control units (i.e., left side of the boundary) and the dotted line the bias-corrected local linear regression for treated units (i.e., right side of the boundary). }
\label{fg:mer1_20}
\end{figure}


\begin{figure}
\centering
\caption{Alternative nonparametric approach - Sharp BRDD $\tau_{SBRD \lvert \mathbb{B}_j}(\mathbf{x})$ (20\% window)}
\includegraphics[width=.5\textwidth]{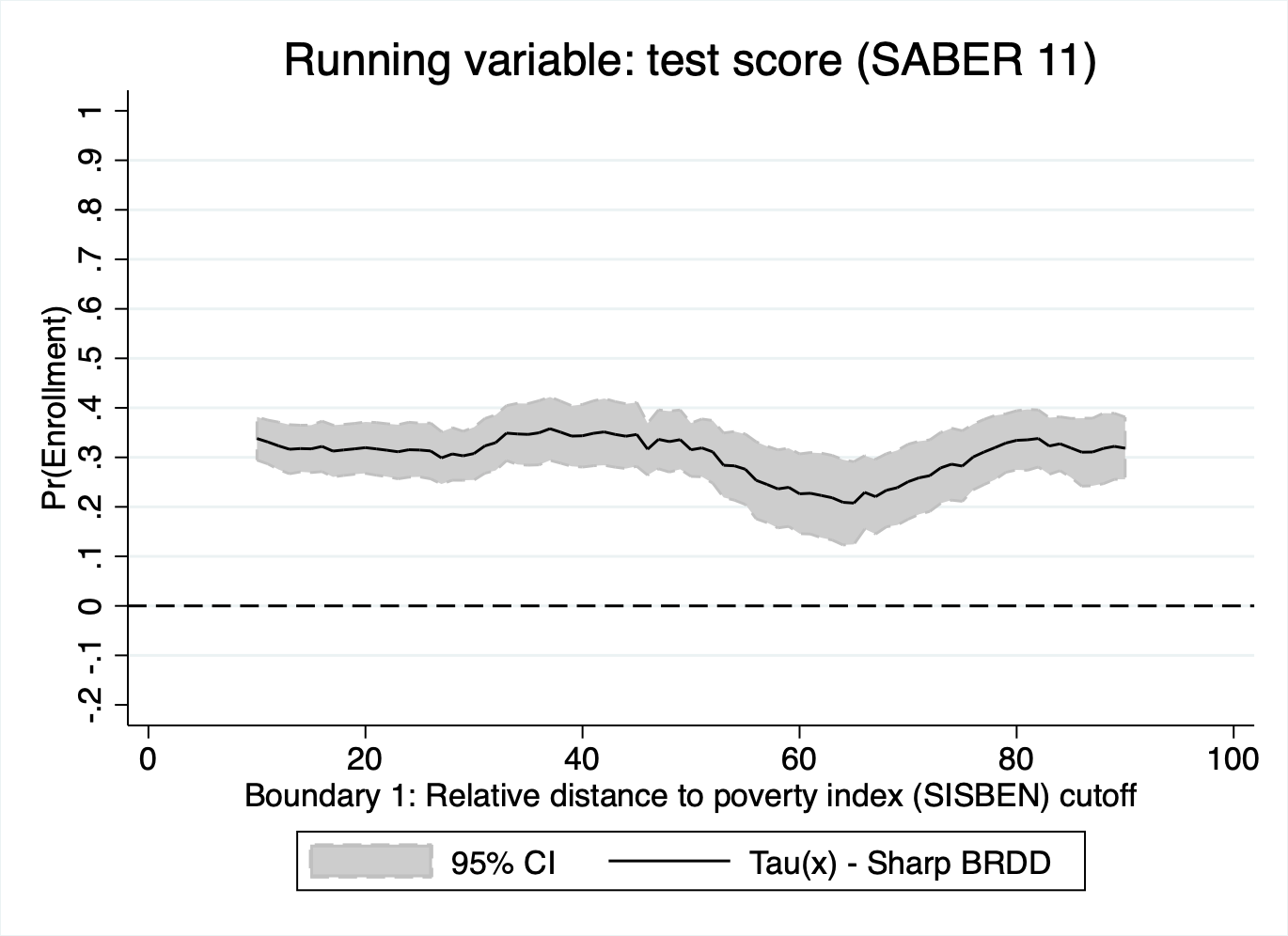}%
\includegraphics[width=.5\textwidth]{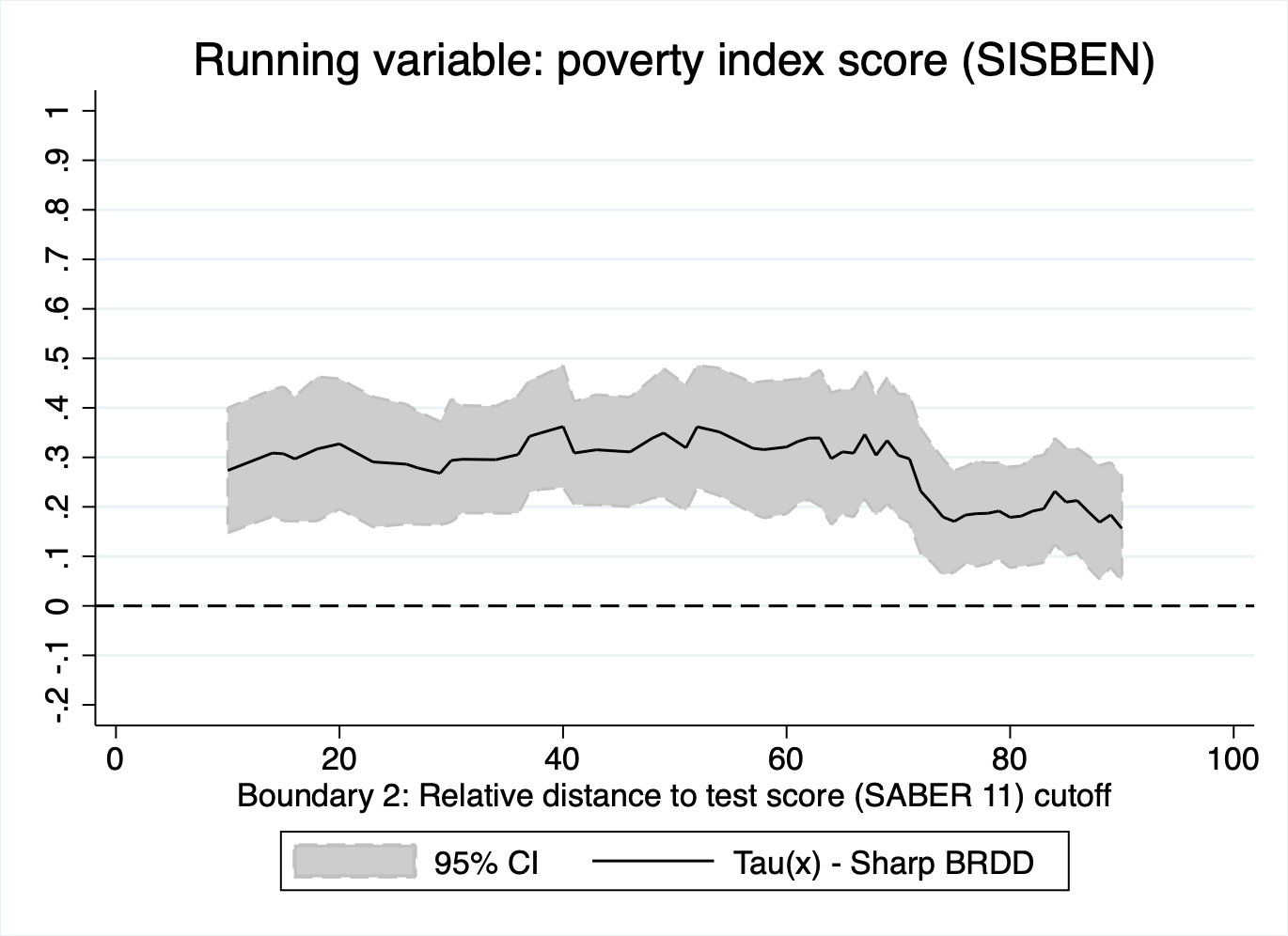}
\caption*{\footnotesize \textbf{Note:} Boundary RD estimates following the flexible nonparametric approach described in Section \ref{mer} and equation (\ref{eq:sbrd_new}) for a moving window that takes 20 percent of the observations at each point along the treatment boundary. The black line plots the bias-corrected local linear regression estimate and the gray area the 95 percent bias-corrected robust confidence interval.}
\label{fg:mer19_20}
\end{figure}

\begin{figure}
\centering
\caption{Alternative nonparametric approach - Fuzzy BRDD $\tau_{FBRD \lvert \mathbb{B}_j}(\mathbf{x})$ (20\% window)}
\includegraphics[width=.5\textwidth]{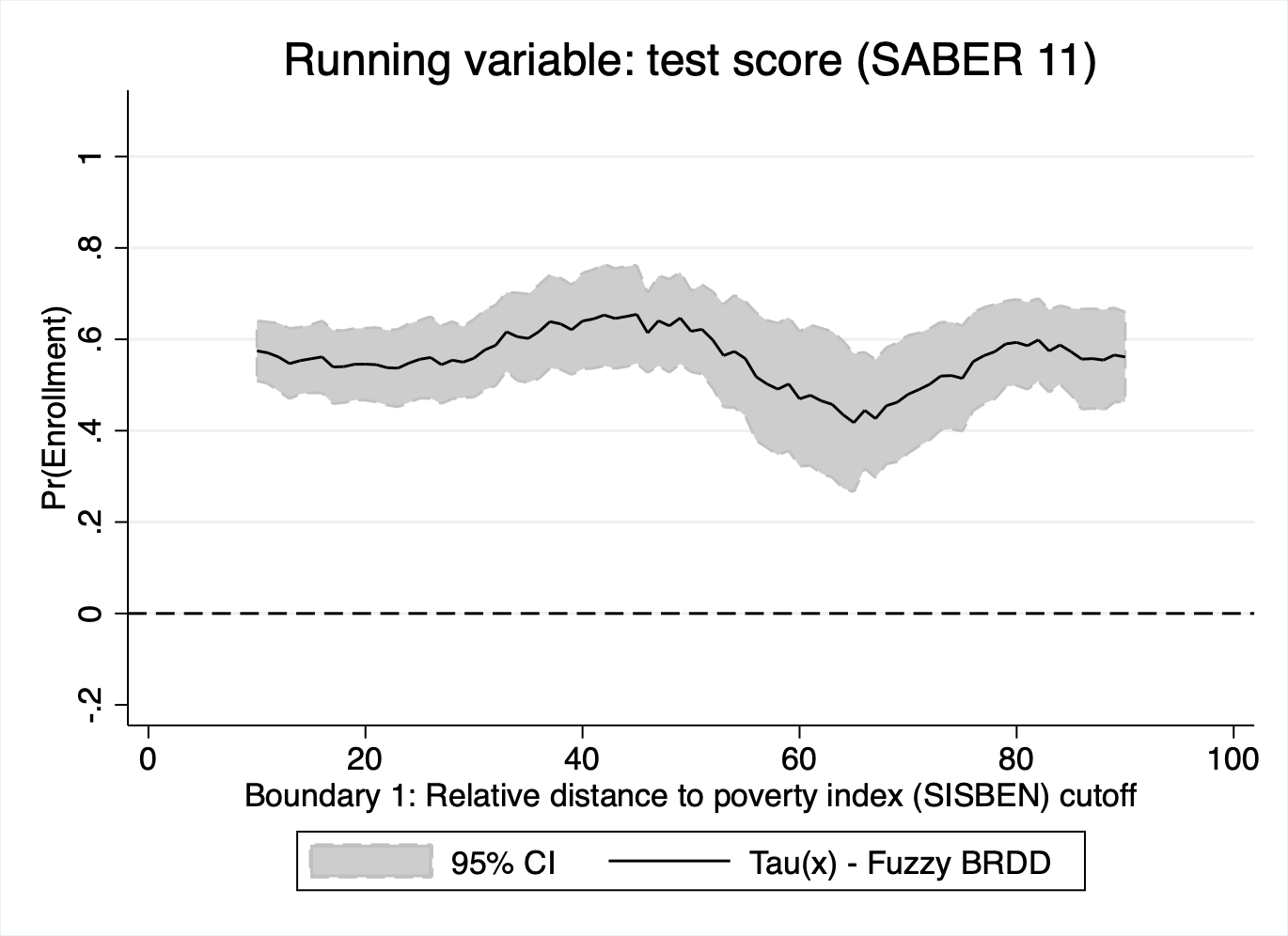}%
\includegraphics[width=.5\textwidth]{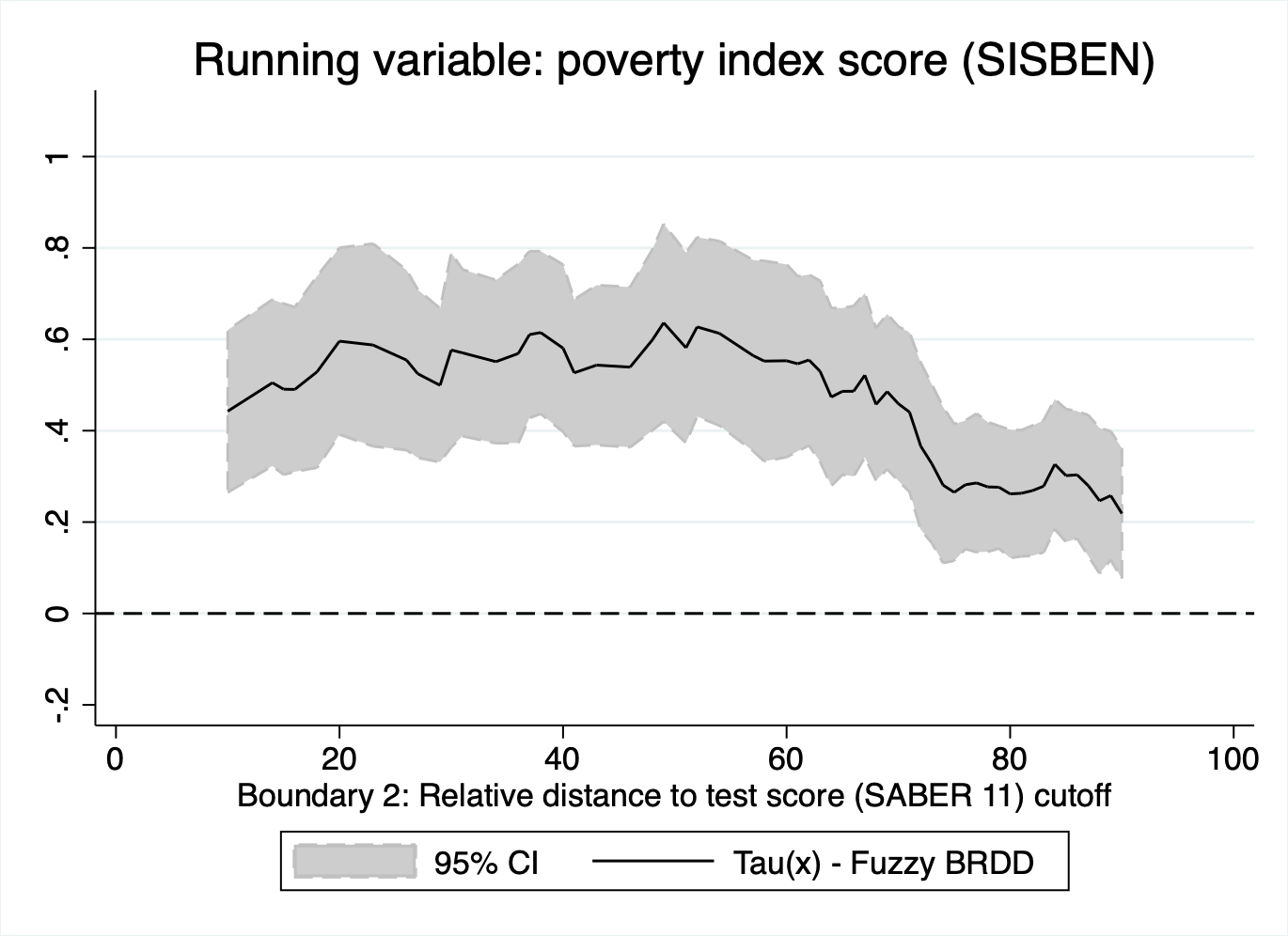}
\caption*{\footnotesize \textbf{Note:} Fuzzy boundary RD estimates following the flexible nonparametric approach described in Section \ref{mer} and equation (\ref{eq:fbrd_new}) for a moving window that takes 20 percent of the observations at each point along the treatment boundary. The black line plots the bias-corrected local linear regression estimate and the gray area the 95 percent bias-corrected robust confidence interval.}
\label{fg:mer19fuzzy_20}
\end{figure}

\begin{figure}
\centering
\caption{Alternative nonparametric approach - Boundary RDD plot (5\% window)}
\includegraphics[width=.5\textwidth]{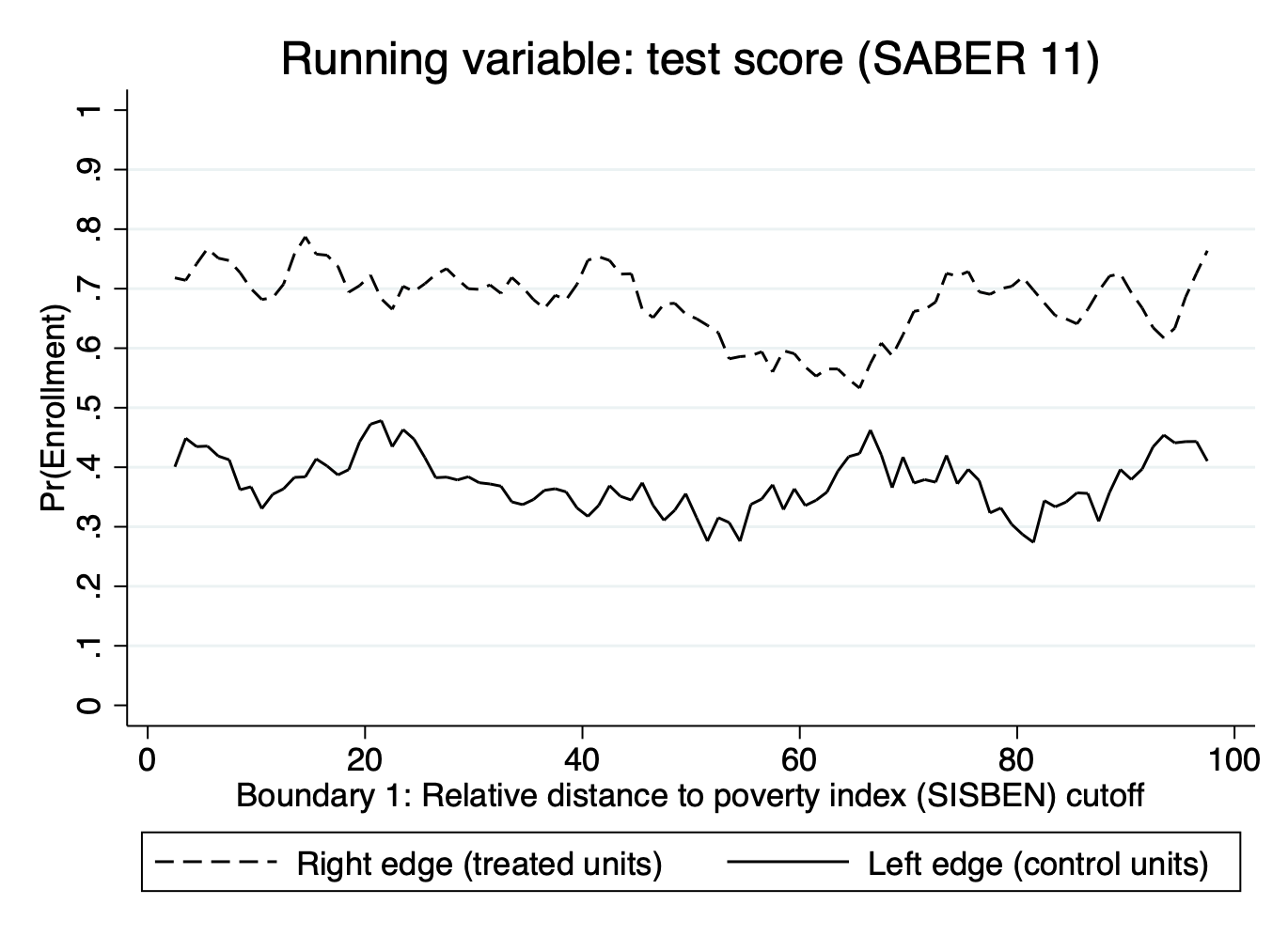}%
\includegraphics[width=.5\textwidth]{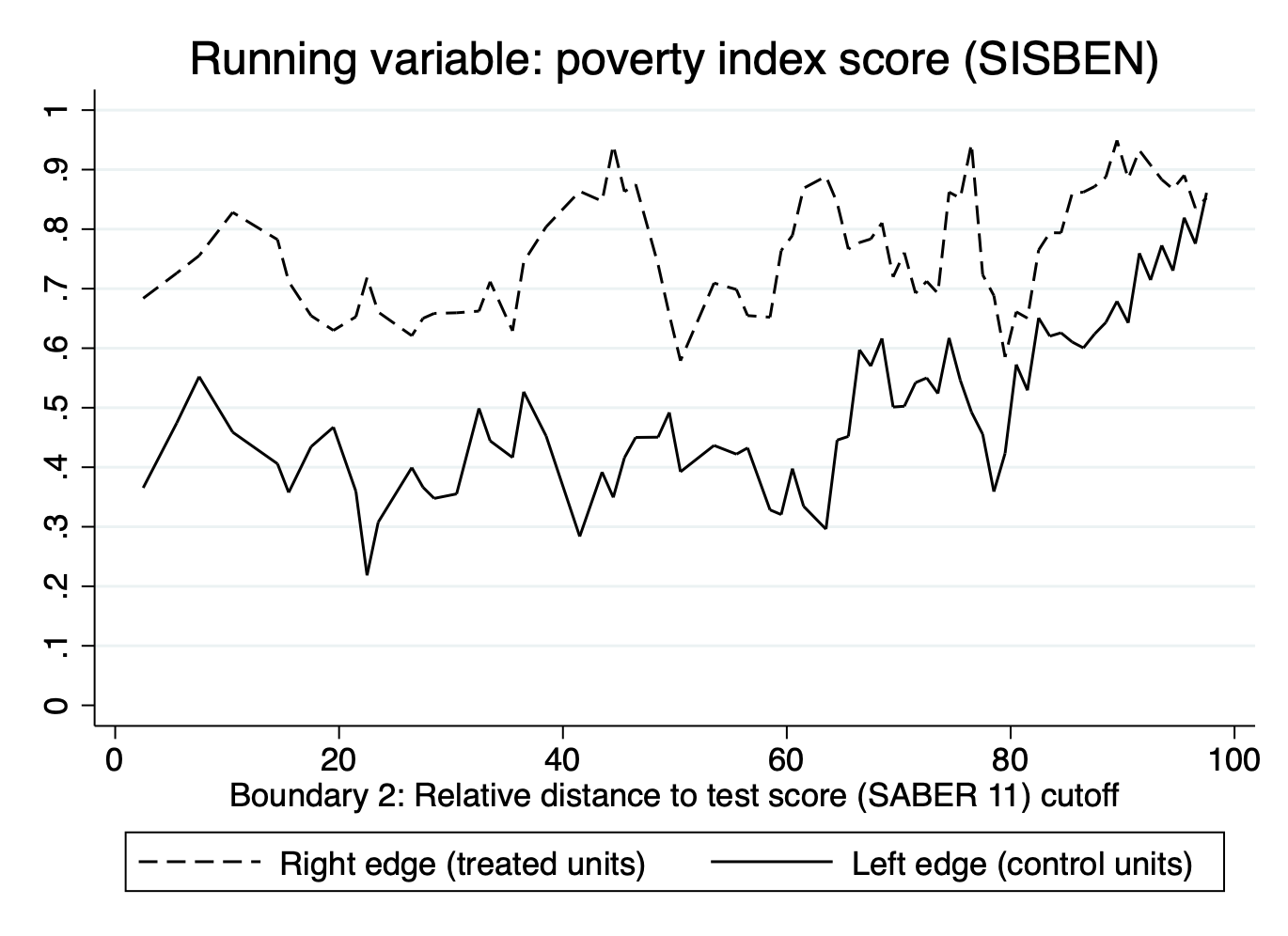}
\caption*{\footnotesize \textbf{Note:} Estimated edges following the flexible nonparametric approach described in Section \ref{mer} for a moving window that takes 5 percent of the observations at each point along the treatment boundary. The bold line plots the bias-corrected local linear regression for control units (i.e., left side of the boundary) and the dotted line the bias-corrected local linear regression for treated units (i.e., right side of the boundary). }
\label{fg:mer1_5}
\end{figure}


%

\begin{figure}[h]
\centering
\caption{Alternative nonparametric approach - Sharp BRDD $\tau_{SBRD \lvert \mathbb{B}_j}(\mathbf{x})$ (5\% window)}
\includegraphics[width=.5\textwidth]{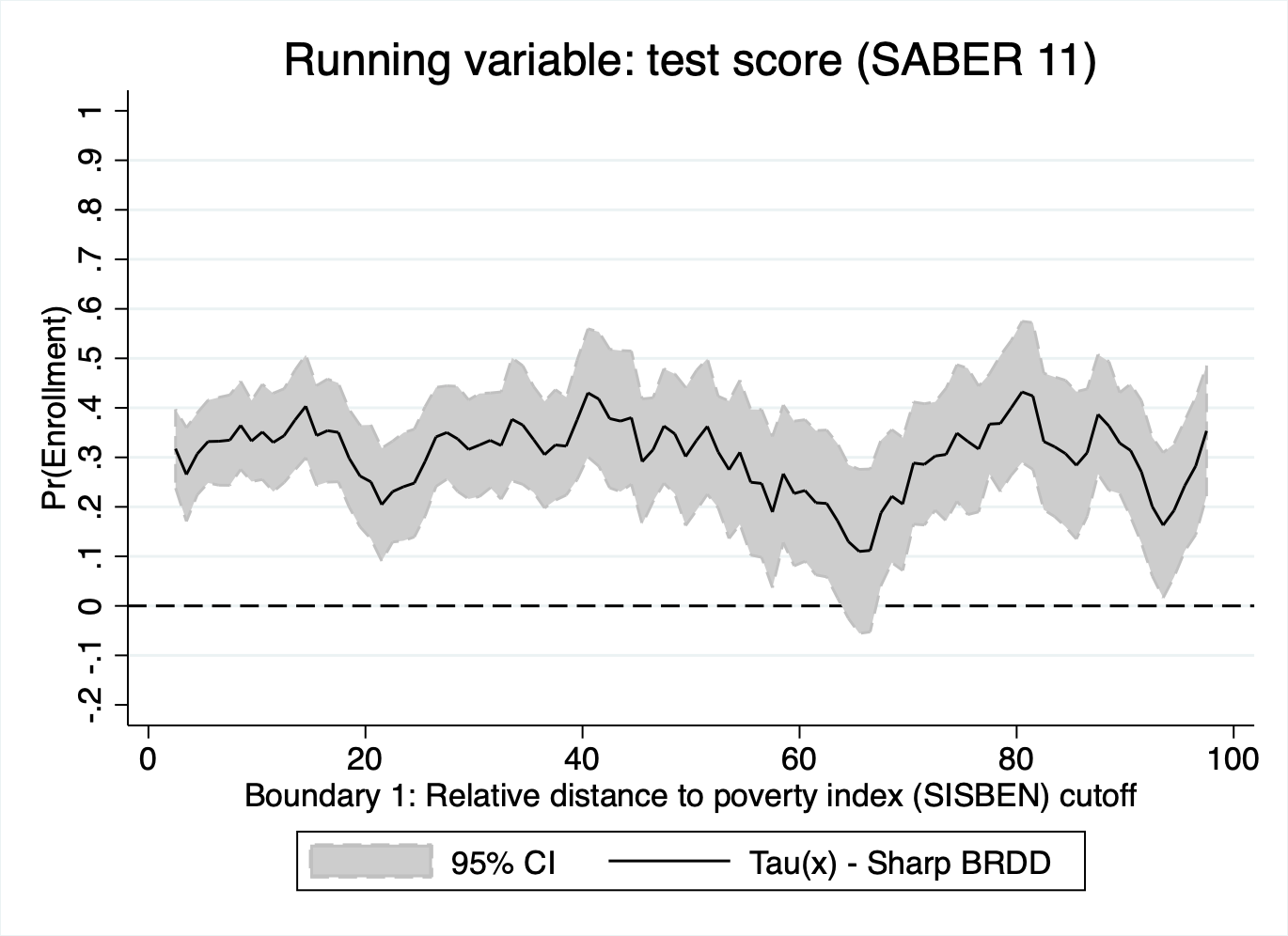}%
\includegraphics[width=.5\textwidth]{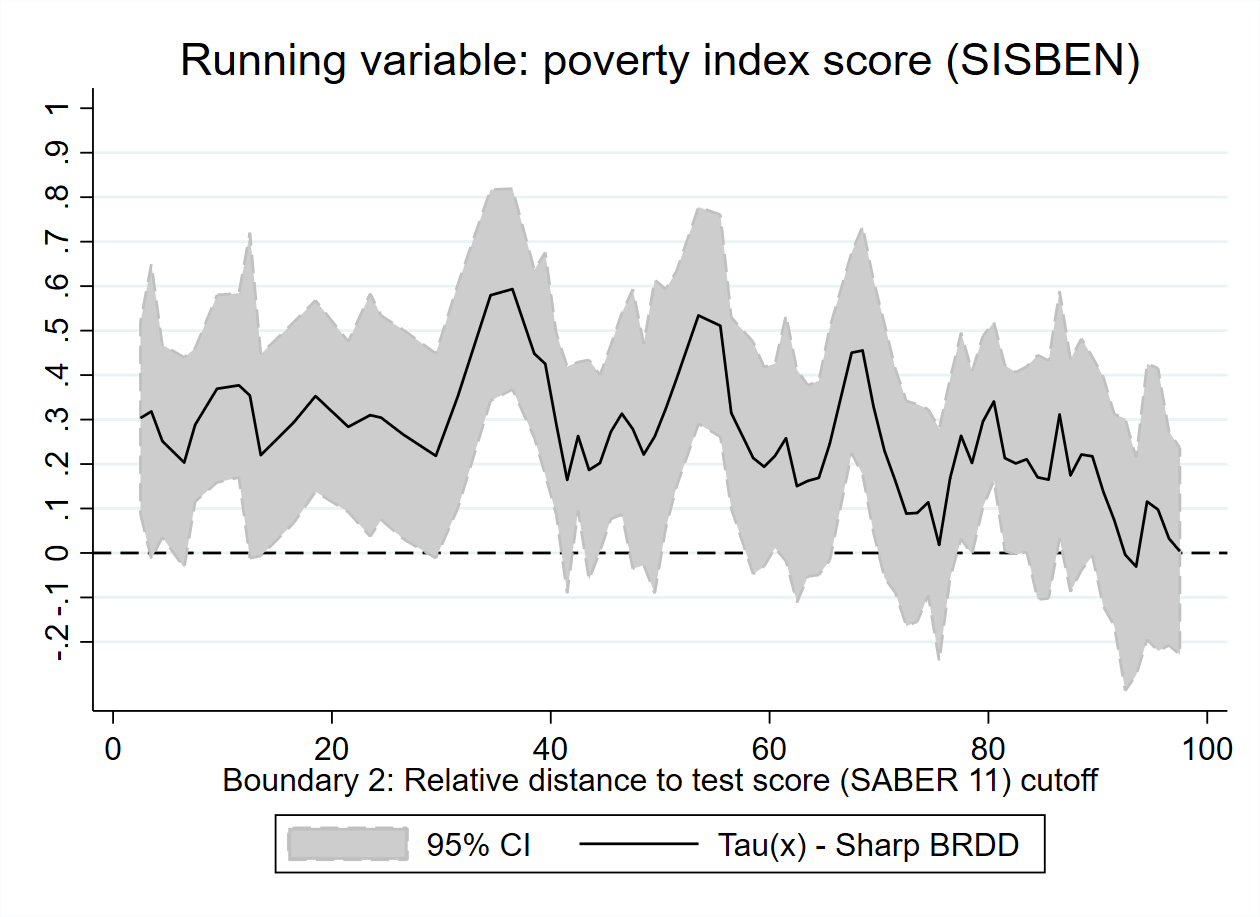}
\caption*{\footnotesize \textbf{Note:} Boundary RD estimates following the flexible nonparametric approach described in Section \ref{mer} and equation (\ref{eq:sbrd_new}) for a moving window that takes 5 percent of the observations at each point along the treatment boundary. The black line plots the bias-corrected local linear regression estimate and the gray area the 95 percent bias-corrected robust confidence interval.}
\label{fg:mer19_5}
\end{figure} \

%

\begin{figure} 
\centering
\caption{Nonparametric approach by \citet{Zajonc2012}}
\includegraphics[width=.5\textwidth]{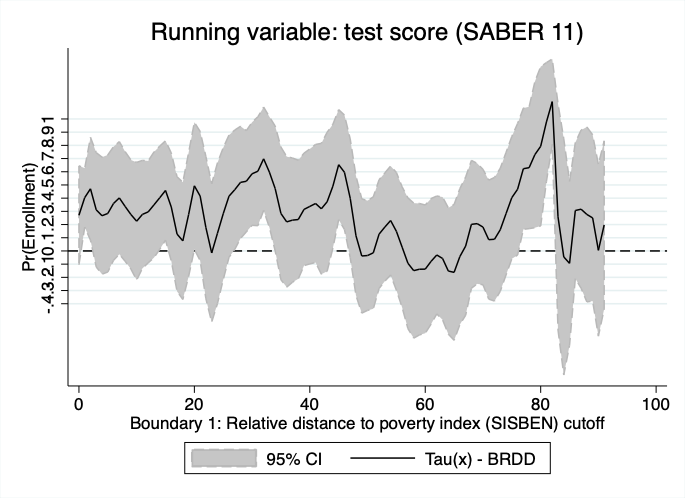}%
\includegraphics[width=.5\textwidth]{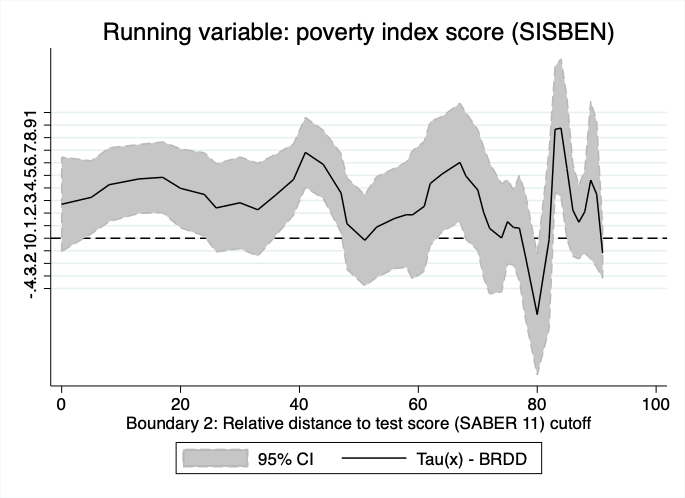}
\caption*{\footnotesize \textbf{Note:}  The black line plots the bivariate local linear regression estimates and the gray area the 95 percent robust confidence interval. The optimal rule-of-thumb bandwidth ($h^*_{ROT}$) is selected as the minimum optimal bandwidth $\hat{h}_{\text{opt}}(\mathbf{x})$ estimated for an evenly spaced grid along the boundary. We drop treatment effects for the top 9 percent of the distributions, as point estimates and confidence intervals become noisy and do not fit the plot correctly. }
\label{fg:zaj12_min}
\end{figure}\

%
\newpage
 \pagebreak[4]
 
\subsection{Supplementary Analysis} \label{sc:app2} 

Table \ref{tab:multicutoffs} estimates a multi-cutoff RD, where we account for the three different scores required for students in main cities, urban, and rural areas. Eligibility scores for the wealth index were 57.21 for students in the country's 14 main cities, 56.32 for other urban areas, and 40.75 for rural areas. These thresholds are equivalent to 42.79, 43.68, and 59.25 when the wealth index scale is inverted.

\begin{table} \centering
\newcolumntype{C}{>{\centering\arraybackslash}X}
\def\sym#1{\ifmmode^{#1}\else\(^{#1}\)\fi}
\caption{\label{tab:multicutoffs} Multi-cutoff estimations for conditional approach}
\begin{tabularx}{\linewidth}{llCCCCC}

\toprule
{}&{$\hat{\tau}$}&{se}&{t-stat}&{p-value}&{ci\_lower}&{ci\_upper} \tabularnewline
\midrule \addlinespace[\belowrulesep]
Main cities (c1=42.79)&.360\sym{***}&.047&7.706&0.00&.269&.452 \tabularnewline
Urban areas (c2=43.68)&.196\sym{***}&.043&4.522&0.00&.111&.282 \tabularnewline
Rural areas (c3=59.25)&.183 &.114&1.590&0.11&-.042&.408 \tabularnewline
Weighted&.258\sym{***}&.030&8.379&0.00&.198&.319 \tabularnewline
Pooled&.274\sym{***}&.027&10.047&0.00&.220&.327 \tabularnewline
\bottomrule
\multicolumn{7}{l}{\footnotesize \sym{*} \(p<0.05\), \sym{**} \(p<0.01\), \sym{***} \(p<0.001\)}\\
\end{tabularx}
\end{table}

We can observe that the treatment effect for students facing the threshold in main cities (36 percentage points) is significantly higher than for those in urban (19.6 percentage points) and rural areas (18.3 percentage points). In the case of the rural areas, the program's effect is in fact not significantly different from zero. We consider that this heterogeneity is relevant from a policy point of view. As we discussed in Section \ref{emp}, the boundary estimation proposed in this paper can easily be extended to the case of multiple cutoffs.

\begin{figure}[h]
\centering
\caption{Multi-cutoff representation for conditional approach}
\includegraphics[width=0.7\textwidth]{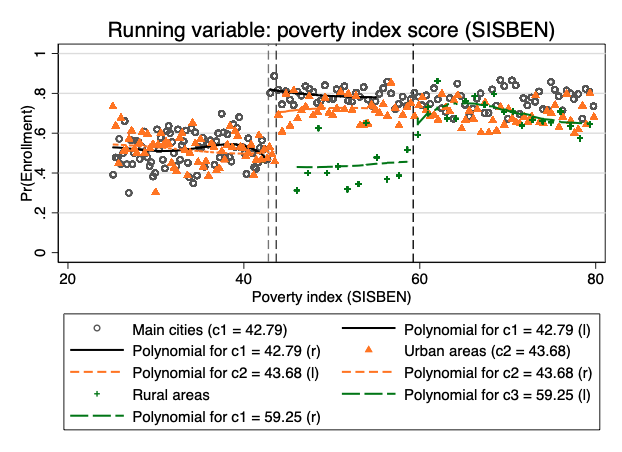}%
\label{fg:multicutoffs}
\end{figure}

Figure \ref{fg:multicutoffs}, illustrates the traditional graphical discontinuity for the multiple thresholds estimation.

\end{spacing}
\end{document}